\documentclass[11pt]{article}

\usepackage[a4paper,margin=1in]{geometry}
\usepackage{amsmath,amssymb,amsthm}
\usepackage{bm}
\usepackage{booktabs}
\usepackage{array}
\usepackage[round,authoryear]{natbib} 
\usepackage[colorlinks=true,linkcolor=blue,citecolor=blue]{hyperref}
\usepackage{xcolor}
\usepackage{graphicx}
\usepackage{relsize}
\usepackage{microtype}             
\usepackage[section]{placeins}   
\graphicspath{{./}}

\newcommand{\safeincludegraphics}[2][width=\textwidth]{%
  \IfFileExists{#2}%
    {\includegraphics[#1]{#2}}%
    {\fbox{\parbox[c][4cm][c]{0.9\textwidth}{\centering
      \texttt{missing: \detokenize{#2}}}}}%
}

\newcommand{\eps}{\varepsilon}
\newcommand{\Pe}{\mathrm{Pe}}
\newcommand{\Bo}{\mathrm{Bo}}
\newcommand{\Ma}{\mathrm{Ma}}
\newcommand{\Rrho}{R_\rho}
\newcommand{\Ct}{\mathcal{C}}         
\newcommand{\Cb}{\mathcal{C}_b}      
\newcommand{\Pdrv}{\mathcal{P}}  
\newcommand{\ubar}{\bar{u}}
\newcommand{\dd}{\,\mathrm{d}}
\newcommand{\code}[1]{\texttt{#1}}
\newcommand{\ctot}{c_{\mathrm{tot}}} 
\newtheorem{remark}{Remark}

\title{Fractionation of polydisperse particles in a receding floating film}
\author{C.\ Choi}
\date{July 18, 2026}

\begin{document}
\maketitle

\begin{center}\small
\end{center}

\begin{abstract}
A thin volatile film carrying $N$ particle species of different sizes
evaporates on a deep immiscible liquid subphase. The spreading
coefficient is positive, so nothing pins. The film ends at a receding
front whose motion falls out of the film equations; no contact-line law
is imposed. We solve the lubrication problem with a conservative
depth-integrated treatment of species transport, and the front turns
out to be a chromatograph. Every species piles into a concentration
spike at the front, each held there in proportion to its P\'eclet
number. The smaller, more diffusive species leaks continuously into the
fluid that survives the front's passage; the larger species is laid
down along the sweep path. The dried deposit is the time-integrated
record of that sweep, and it is sorted by size: small-rich centre,
large-rich mid-annulus. Pinned bidisperse droplets sort the other way.
Depinned droplets on solids sort this way, but by a force balance;
here, diffusivity contrast alone picks the direction. Size enters the
transport problem only through the P\'eclet number, so diffusivity
contrast is the only symmetry-breaking available to choose a direction:
Marangoni stresses, colloidal interactions and wetting effects act on
the magnitude and, by the symmetry of the transport operator, cannot
set the direction. Adjacent-species band separations scale with the
difference in inverse effective P\'eclet number, a law derived and
measured here: the resolution of the chromatograph. That scaling points
to a route to size fractionation of polydisperse nanoparticles in the
few-nanometre regime, exactly where standard methods struggle.
\end{abstract}

\section{Introduction}
\label{sec:intro}

A particle-laden droplet does not dry uniformly. The canonical example
is the coffee stain: capillary flow towards the pinned contact line
carries particles outward and dumps them at the edge
\citep{deegan1997,deegan2000}. The deposit morphology can be steered:
thermal and solutal Marangoni stresses, substrate patterning, the
evaporation profile \citep{hu2006,karapetsas2016}. That control is
exploited hard in evaporative self-assembly, inkjet printing and
biochemical microarrays.

What is far less understood is how a polydisperse population gets
sorted by size while the deposit forms. Colloidal and nanoparticle
syntheses are polydisperse whether we like it or not. Fractionating
them in the few-nanometre regime remains hard. Transport there is
diffusion, not sedimentation or filtration, and the standard methods
lean on the latter two.

Size sorting during evaporation has been seen before, and sometimes
named for what it is. \citet{wong2011} demonstrated a
`nanochromatography' driven by the coffee-ring effect: particles are
selected at the contact line by matching their diameter to the local
meniscus thickness. The picture has since been formalized in the
language of classical chromatography \citep{liamtsau2022}. For
bidisperse suspensions on a substrate the established ordering is
small-at-edge, large-towards-centre \citep{patil2018}. Pinned-droplet
studies reproduce that direction and explain it through contact-line
capture and differential radial transport \citep{zolotarev2022}.
Recession alters the deposit even without polydispersity:
\citet{freedbrown2014} showed that a smoothly receding line on a solid
replaces the edge ring with a centre-heavy deposit.
Depinning on a solid reverses the bidisperse ordering to
small-centre/large-ring \citep{iqbal2018}. The mechanism there is a
drag-based force balance in a high-contact-angle cap with no wedge.

Essentially all of this work rests on two structural assumptions: a
solid substrate and a pinned contact line.
Recent asymptotic treatments also work in a regime where radial
diffusion is negligible at leading order \citep{dambrosio2026}. In that
limit the deposit is size-blind at the level of transport, by
construction.

The unpinned, substrate-free floating film has not yet been studied
and it is physically different. A volatile film floating on a deep
immiscible subphase with positive spreading coefficient has no triple
line to pin and no solid to impose a boundary condition. The film ends
at a freely receding contact line whose motion comes out of the film
equations themselves. Nothing guarantees that the sorting behaviour
established for pinned deposits on solids carries over to this
geometry. It does not.

Here we derive and solve an $N$-species lubrication model for a thin
volatile film evaporating on a deep liquid subphase. Size is
the only property that distinguishes the species, and it enters the
transport problem through the Stokes--Einstein diffusivity and nowhere
else. The receding front concentrates all species into a common spike
and retains each in proportion to its P\'eclet number. The deposit is
the time-integrated record of this sweep: radially sorted, small-rich
centre, large-rich mid-annulus, opposite to the pinned-substrate
ordering. The species differ in nothing but their P\'eclet numbers;
whatever sorting the deposit shows can have come from nowhere else.
It needs no Marangoni stresses and no particle forces.
Those modify the magnitude of the effect and leave its direction alone,
by the symmetry of the species operator (\S\ref{sec:theorem}); we
verify that invariance directly for the evaporation closure, the
crowding rheology, the mobility gate and the precursor floor, and rest
the remainder on the theorem.
Adjacent-species band separations scale with the difference
in inverse effective P\'eclet number, a law we derive from an expansion
in diffusivity, confirm numerically, and read as the resolution
of the mechanism.

The paper is organized as follows. Section~\ref{sec:formulation}
formulates the $N$-species evaporating-film model and its lubrication
reduction; section~\ref{sec:results} presents the frontal-fractionation
mechanism and the deposits it writes; section~\ref{sec:discussion}
closes with the regime of validity, the contrast with pinned-substrate
deposition, and the consequences for nanoparticle fractionation.

\section{Formulation}
\label{sec:formulation}

\emph{Conventions.} Hats denote dimensional quantities; unhatted symbols
are dimensionless. Cylindrical coordinates $(\hat r,\hat z)$,
axisymmetric, no swirl. The film is decane, the subphase ethylene glycol
(EG), and the particles rigid spheres of radii $\hat a_i$,
$i=1,\dots,N$. Code symbols from
\code{multi\_size\_qd\_v14\_soft\_Nsize\_v4p.py} appear in
typewriter font; Appendix~\ref{sec:dictionary} collects the full mapping
table.

\subsection{Geometry, configuration, and standing assumptions}
\label{sec:geometry}

A volatile liquid film occupies $\hat s(\hat r,\hat t)<\hat z<
\hat\eta(\hat r,\hat t)$, where $\hat z=\hat s$ is the (deformable)
decane--EG interface and $\hat z=\hat\eta=\hat s+\hat h$ the decane--gas
interface, with $\hat h$ the local film thickness. The film is deposited
as a cap of initial centre thickness $\hat H_0$ and lateral extent
$\hat L_0$, and it spreads: the spreading coefficient
\begin{equation}
\hat S \;=\; \hat\gamma_{ea}-\hat\gamma_{da}-\hat\gamma_{de}
\;=\; +12.2\ \mathrm{mN\,m^{-1}}\;>\;0,
\label{eq:spreading}
\end{equation}
where $\hat\gamma_{ea}=47.7\ \mathrm{mN\,m^{-1}}$ (EG--air) and
$\hat\gamma_{da}=23.8\ \mathrm{mN\,m^{-1}}$ (decane--air) are measured
directly and $\hat S$ is measured independently. The decane--EG tension
is therefore the \emph{inferred} member of the triple,
$\hat\gamma_{de}=\hat\gamma_{ea}-\hat\gamma_{da}-\hat S
=11.7\ \mathrm{mN\,m^{-1}}$. Since $\hat S>0$
there is no Neumann lens and no pinned contact line. The film is bounded
by a freely moving apparent contact line, ahead of which lies a thin
adsorbed (precursor) film maintained by the disjoining/evaporation
balance of \S\ref{sec:evap}. This is the defining geometric difference
from the sessile-droplet literature
\citep{dambrosio2026,malachtari2025}: there, contact-line kinematics
are imposed (CR/CA/SS/SJ) or mediated by a solid; here the motion of the
front is \emph{emergent} from the film equations.

Standing assumptions, each revisited where it enters:
\begin{itemize}
\item[\textbf{A1}] \textbf{(thinness)} $\eps=\hat H_0/\hat L_0\ll 1$.
Physically $\hat H_0\sim 10^2\,$nm--$1\,\mu$m and $\hat L_0\sim 1\,$mm, so
$\eps\sim10^{-4}$--$10^{-3}$.
\item[\textbf{A2}] \textbf{(Newtonian, incompressible, isothermal)} Both
liquids are Newtonian and incompressible. Isothermal baseline: thermal
Marangoni stresses enter only through the retired \code{MAR} toggle
(\S\ref{sec:filmeq}); the paper's direction claim is established at
$\code{MAR}=0$ and extends to $\code{MAR}\neq0$ by \S\ref{sec:theorem},
since $\sigma$ is a collective field.
\item[\textbf{A3}] \textbf{(viscosity ratio)}
$\hat\mu_e/\hat\mu_d\approx 16.1/0.86\approx 19\gg1$: the subphase is far
more viscous than the film. The large ratio justifies the no-slip condition
at $\hat z=\hat s$ for the \emph{film} problem (\S\ref{sec:profile}). The
subphase nonetheless responds quasi-statically: its capillary-gravity
relaxation time, though lengthened by $\hat\mu_e$, remains
$10^{-4}$--$10^{-2}$ of the drying time (Appendix~\ref{sec:relax}).
\item[\textbf{A4}] \textbf{(density ratio)}
$\Rrho=\hat\rho_d/\hat\rho_e=730/1113=0.656$. This is the code constant
$\code{Rrho}=0.66$: the measured decane/EG density ratio, not a fitted
value.
\item[\textbf{A5}] \textbf{(dilute-to-moderate loading)} Particle volume
fraction enters the momentum balance only through an effective viscosity
$\hat\mu(\ctot)$ (\S\ref{sec:rheology}) and the transport problem through a
mobility gate (\S\ref{sec:gate}). Stokes--Einstein diffusion is retained
at leading order. That retention is the entire point (\S\ref{sec:species},
\S\ref{sec:front}).
\item[\textbf{A6}] \textbf{(passive gas)} $\hat\mu_g/\hat\mu_d\ll1$; the gas
exerts no traction and enters only through the evaporative flux closure
(\S\ref{sec:evap}).
\end{itemize}

\subsection{Governing equations (dimensional)}
\label{sec:governing}

In each liquid ($k=d$ for the film, $k=e$ for the subphase):
\begin{equation}
\hat\nabla\cdot\hat{\bm u}_k = 0, \qquad
\hat\rho_k\!\left(\partial_{\hat t}\hat{\bm u}_k
 + \hat{\bm u}_k\cdot\hat\nabla\hat{\bm u}_k\right)
 = -\hat\nabla\hat p_k
 + \hat\nabla\cdot\!\left[\hat\mu_k\!\left(\hat\nabla\hat{\bm u}_k
 + \hat\nabla\hat{\bm u}_k^{\!\top}\right)\right]
 - \hat\rho_k\hat g\,\bm e_z.
\label{eq:NS}
\end{equation}
In the film, $\hat\mu_d=\hat\mu_{d,0}\,\mu(\ctot)$ with
$\ctot=\sum_i c_i$ the total scaled particle concentration
(\S\ref{sec:rheology}). Each species obeys an advection--diffusion
equation
\begin{equation}
\partial_{\hat t}\hat c_i+\hat{\bm u}_d\cdot\hat\nabla\hat c_i
 = \hat\nabla\cdot\!\left(\hat D_i\hat\nabla\hat c_i\right),
\qquad
\hat D_i=\frac{k_B\hat T}{6\pi\hat\mu_{d,0}\,\hat a_i},
\label{eq:AD}
\end{equation}
with \textbf{no-flux conditions at both interfaces} (particles neither
evaporate nor, in the baseline, adsorb):
\begin{equation}
\left[\hat D_i\hat\nabla\hat c_i
 - \hat c_i\big(\hat{\bm u}_d-\hat{\bm u}_\Sigma\big)\right]\!\cdot\bm n = 0
\quad\text{on } \hat z=\hat s \text{ and } \hat z=\hat\eta,
\label{eq:noflux}
\end{equation}
where $\hat{\bm u}_\Sigma$ is the interface velocity. At the top
interface this includes the recession due to evaporation. That is what
lets evaporation concentrate particles \emph{without ever appearing in
the species equation} (\S\ref{sec:depthint}). Stokes--Einstein is the
only place particle size enters the transport problem:
$\hat D_i\propto1/\hat a_i$, so an $N$-size distribution is an
$N$-diffusivity distribution. $\hat D_i$ is evaluated at the solvent
viscosity $\hat\mu_{d,0}$. A crowding-corrected
$\hat D_i\propto1/\mu(\ctot)$ would rescale every diffusivity by a
common, size-blind factor, preserving the $\Pe$ ratios that set the
ordering of \S\ref{sec:front}. It therefore belongs to the magnitude
channel, alongside the hindered-diffusivity discussion of
\S\ref{sec:gate}. Interfacial trapping (Pickering) and steric exclusion
are higher rungs of the activation ladder and are omitted from the
baseline derivation; they modify the magnitude of the segregation
without changing its sign (\S\ref{sec:theorem}, \S\ref{sec:expansion}).

\subsection{Interfacial conditions (dimensional)}
\label{sec:interface}

\textbf{Decane--gas interface} $\hat z=\hat\eta$, outward normal $\bm n$,
evaporative mass flux $\hat J$ (mass/area/time):
\begin{itemize}
\item Kinematic $+$ phase change:\quad
$\hat\rho_d(\hat{\bm u}_d-\hat{\bm u}_\Sigma)\cdot\bm n=\hat J$.
\item Normal stress:\quad
$\bm n\cdot\hat{\bm T}_d\cdot\bm n
 = -\hat p_g-\hat\gamma_{da}\,2\hat\kappa-\hat\Pi(\hat h)$,
where $2\hat\kappa$ is the mean curvature of $\hat z=\hat\eta$ and
$\hat\Pi$ the disjoining pressure of the \emph{film}: for van der Waals
interaction across a film of thickness $\hat h$,
$\hat\Pi(\hat h)=-\hat A_H/(6\pi\hat h^3)$ with Hamaker constant
$\hat A_H>0$ (attractive; the sign follows from the evaporation closure
of \S\ref{sec:evap} and is not adjustable).
\item Tangential stress:\quad
$\bm t\cdot\hat{\bm T}_d\cdot\bm n=\hat\nabla_s\hat\sigma$, with
$\hat\sigma$ the surface tension; at baseline $\hat\nabla_s\hat\sigma=0$.
\item Vapour recoil and thermal effects: neglected (A2, A6).
\end{itemize}

\textbf{Decane--EG interface} $\hat z=\hat s$: continuity of velocity and
traction, plus the interfacial tension $\hat\gamma_{de}$ in the normal
balance. In the lubrication limit with A3, these reduce for the
\emph{film} to no-slip at $\hat z=\hat s$ (corrections are
$O(\hat\mu_d/\hat\mu_e)\approx0.05$, i.e.\ an effective Navier slip length
$\hat b\sim(\hat\mu_d/\hat\mu_e)\hat h$; retained as a remark, dropped at
leading order), and for the \emph{subphase} to a normal-load balance
(Appendix~\ref{sec:subphase}).

\textbf{Far field} $\hat r\to\hat L_x$: flat precursor film in evaporative
equilibrium (\S\ref{sec:evap}), $\hat s\to0$, no flux.

\subsection{Scaling and nondimensionalization}
\label{sec:scaling}

\begin{equation}
\hat r=\hat L_0\,r,\quad
(\hat z,\hat h,\hat s,\hat\eta)=\hat H_0\,(z,h,s,\eta),\quad
\hat u_r=\hat U u,\quad
\hat u_z=\eps\hat U w,\quad
\hat t=\frac{\hat L_0}{\hat U}\,t,
\end{equation}
\begin{equation}
(\hat p,\hat\Pi)=\frac{\hat\mu_{d,0}\hat U}{\eps^2\hat L_0}\,(p^{*},\Pi),
\qquad
\hat c_i=\hat c_{\mathrm{ref}}\,c_i,
\qquad
\hat J=\hat J_{\mathrm{ref}}\,J.
\end{equation}
The velocity scale $\hat U$ is left free (fixed by the choice of the
dimensionless surface-tension group below); the reduced Reynolds number
$\hat\rho_d\hat U\hat H_0\eps/\hat\mu_d\ll1$, so inertia drops out. The
dimensionless groups, with their code names:

\begin{center}
\small
\begin{tabular}{@{}lllll@{}}
\toprule
Group & Definition & Code & Value & Anchor \\
\midrule
$\eps$ & $\hat H_0/\hat L_0$ & n/a & $10^{-4}$--$10^{-3}$ & geometry \\
$\Ct$ & $\eps^3\hat\gamma_{da}/(\hat\mu_{d,0}\hat U)$ & \code{gam\_t} & $0.01$ & fixes $\hat U$ \\
$\Cb$ & $(\hat\gamma_{de}/\hat\gamma_{da})\,\Ct$ & \code{gam\_b} & $0.0049$ & $\hat\gamma_{de}=11.7$ from $\hat S$ \\
$\Bo$ & film weight / driving pressure & \code{Bo} & $0.05$ & gravity \\
$\Rrho$ & $\hat\rho_d/\hat\rho_e$ & \code{Rrho} & $0.66$ & $=730/1113$, physical \\
$k_s$ & $\Bo/\Rrho$ & \code{k\_sub} & $0.076$ & Archimedes, App.~\ref{sec:winkler}; derived \\
$E$ & $t_{\mathrm{cap}}/t_{\mathrm{evap}}$ & \code{E} & $0.005$ & slow evaporation \\
$\Delta$ & Kelvin (pressure) sensitivity & \code{Delta} & $10^{-3}$ & \S\ref{sec:evap} \\
$K$ & conductive/kinetic resistance ratio & \code{K} & $0.1$ & \S\ref{sec:evap} \\
$A$ & scaled Hamaker group & \code{A} & $10^{-6}$ & \S\ref{sec:evap}; tied to $h_p$ \\
$\Pe_i$ & $\hat U\hat L_0/\hat D_i$ & \code{PE[i]} & $10^4$--$1.6{\times}10^5$ & $\Pe_i\propto\hat a_i$ \\
$\tau$ & subphase relaxation / advective time & \code{tau\_sub} & $10^{-2}$ & anchored, App.~\ref{sec:relax} \\
\bottomrule
\end{tabular}
\end{center}

Two consistency remarks. (i) $\code{Rrho}=0.66$ and $\code{gam\_b}=0.0049$
are both anchored to measurement: the first is the decane/EG density
ratio to two digits, and the second follows from the measured spreading
coefficient. Neither is tuned.
(ii) The particle P\'eclet regime is
\begin{equation}
\eps^2\Pe_i\;\ll\;1\;\ll\;\Pe_i
\label{eq:regime}
\end{equation}
with $\eps^2\Pe_i\lesssim1.6\times10^{5}\times10^{-6}\approx0.16$ at the
upper end of the physical $\eps$ range, and smaller by up to four orders
of magnitude at $\eps=10^{-4}$. Vertical diffusion equilibrates the
concentration across
the film (left inequality, \S\ref{sec:vertical}). Radial diffusion is
weak \emph{in the bulk} (right inequality) but is promoted to leading
order inside the front boundary layer, whose width self-selects to make
it so (\S\ref{sec:front}). This is precisely the corner excluded by
construction in \citet{dambrosio2026}: their regime
$\theta_0^2\ll\theta_0^2\Pe\ll1$ removes diffusion from the leading-order
radial transport entirely. In their limit the deposit is size-blind at
the level of transport; in ours the front sorts by size. The two
statements describe different distinguished limits, they do not
conflict.

\subsection{Leading-order film problem}
\label{sec:film}

\subsubsection{Velocity profile and flux}
\label{sec:profile}

At $O(1)$ the radial momentum equation and continuity give the standard
lubrication balance in the film, with the driving pressure $\Pdrv(r,t)$
independent of $z$:
\begin{equation}
\partial_z\!\left(\mu(\ctot)\,\partial_z u\right)=\partial_r\Pdrv,
\qquad
\Pdrv \;=\; p+\Bo\,(h+s),
\label{eq:lub}
\end{equation}
where $p$ is the capillary/disjoining part (\S\ref{sec:pressure}) and
$\Bo(h+s)$ the hydrostatic head measured from the \emph{deformed} free
surface. The subphase deflection $s$ enters the driving pressure here,
and this is the only place softness couples back to the flow. Boundary
conditions: no-slip at the liquid--liquid interface (A3), Marangoni-loaded
shear at the top:
\begin{equation}
u=0\ \text{at } z=s,
\qquad
\mu\,\partial_z u=\partial_r\sigma\ \text{at } z=\eta.
\end{equation}
Solving,
\begin{equation}
u(z)=\frac{1}{\mu}\,\partial_r\Pdrv\left(\frac{\zeta^2}{2}-h\zeta\right)
 +\frac{\zeta}{\mu}\,\partial_r\sigma,
\qquad \zeta\equiv z-s\in[0,h],
\end{equation}
and the volume flux and depth-averaged velocity are
\begin{equation}
Q=\int_0^h u\dd\zeta
 = -\frac{h^3}{3\mu}\,\partial_r\Pdrv+\frac{h^2}{2\mu}\,\partial_r\sigma,
\qquad
\ubar=\frac{Q}{h}.
\label{eq:flux}
\end{equation}
These are exactly the code's flux terms (\code{h\_pos**3/3*grad(Phi)},
\code{h\_pos**2/2*grad(sigma)} with \code{Phi = p + Bo*eta\_top}). The
$O(\hat\mu_d/\hat\mu_e)$ slip correction would add $\hat b\,h^2$-type
mobility ($\hat b\approx0.05\,h$). It uniformly \emph{increases} mobility
and never touches the structure of the species operator. This belongs
to the magnitude channel, and we drop it at leading order.

\subsubsection{Crowding rheology}
\label{sec:rheology}

$\mu(\ctot)$ follows a Krieger--Dougherty-type law regularized for Newton
robustness,
\begin{equation}
\mu=\big(1-\Phi_{\max}\tanh(\chi\,\ctot/\Phi_{\max})\big)^{-2},
\qquad \Phi_{\max}=0.95,\quad \chi=0.01,
\label{eq:KD}
\end{equation}
which recovers the Krieger--Dougherty form at small argument and
saturates, rather than diverging, at extreme spike amplitudes; same
physics lineage as \citet{karapetsas2016} and \citet{malachtari2025}.
The correction is weak through the bulk, where $\ctot=O(1$--$5)$ gives
$\mu\approx1.0$--$1.1$, but it is \emph{not} weak at the front: at the
spike amplitudes of figure~\ref{fig:spike}(a), \eqref{eq:KD} returns
$\mu$ of order $10$, rising further in the final collapse and bounded
above by the saturation value $(1-\Phi_{\max})^{-2}=400$ that the
$\tanh$ regularization imposes in place of a divergence. The crowding
term is retained in the operator precisely so that the direction claim
is established in its presence rather than in its absence. Its
size-blindness ($\mu$ sees only the collective $\ctot=\sum_i c_i$)
guarantees it cannot set the sorting direction in any regime; it acts
on the magnitude channel alone.

\subsubsection{Film evolution equation}
\label{sec:filmeq}

Mass conservation of the film with evaporative loss:
\begin{equation}
\partial_t h+\frac{1}{r}\,\partial_r\!\left[r\!\left(
 -\frac{h^3}{3\mu}\partial_r\Pdrv+\frac{h^2}{2\mu}\partial_r\sigma
\right)\right] = -E\,J
\tag{F}
\label{eq:F}
\end{equation}
with $\sigma=1-\Ma\,T_s$ reducing to $\sigma\equiv1$ at baseline
($\code{MAR}=0$). Equation \eqref{eq:F} is exactly the film residual
solved by the code.

\subsection{Curvature/pressure equation and the mixed formulation}
\label{sec:pressure}

The leading-order normal stress balance at $z=\eta$ gives
\begin{equation}
p \;=\; \Pi(h)\;-\;\Ct\,\frac{1}{r}\,\partial_r\!\big(r\,\partial_r\eta\big),
\qquad
\Pi(h)=-\frac{A}{h^3},
\qquad
\eta=h+s.
\tag{P}
\label{eq:P}
\end{equation}
The curvature acts on the \textbf{total elevation} $\eta=h+s$: the
capillary restoring force sees the actual liquid surface riding on the
deformed subphase, not the film thickness alone. Equations
\eqref{eq:F}$+$\eqref{eq:P} are fourth order in $h$; the code solves
them as the mixed pair $\{h,p\}$ with $C^2$ elements. That is a
discretization choice, not a modelling one. It is recorded in
Appendix~\ref{sec:benchmark}.
Equation \eqref{eq:P} is also where the
$h_{\mathrm{pos}}$ regularization
$h\mapsto\tfrac12\big(h+\sqrt{h^2+4h_{\mathrm{reg}}^2}\big)$ enters every
division: a smooth positive part whose numerical smoothing scale
$h_{\mathrm{reg}}=h_p/10$ sits strictly below the physical precursor
scale. The two scales are never conflated. With that separation the
regularized Kelvin identity of \S\ref{sec:evap} holds to
$O\big((h_{\mathrm{reg}}/h_p)^2\big)$: the exact $J=0$ equilibrium sits
at $h=h_p\big(1-(h_{\mathrm{reg}}/h_p)^2\big)=0.99\,h_p$, and a flat film
initialized at $h=h_p$ carries a residual flux $J\approx0.03$ that
decays within the first time units. (Conflating $h_{\mathrm{reg}}$ with
$h_p$, the natural first implementation, displaces the equilibrium
to $h\to0$ and leaves the nominal precursor evaporating at $0.76$ of
full drive. Recorded here as a second correctness trap alongside
Remark~\ref{rem:trap}.) Unlike an algebraic clamp, the regularization
leaves the \emph{state} untouched. The film cannot go negative, because
\S\ref{sec:evap} makes the precursor a true equilibrium.

\subsection{Evaporation closure and the Kelvin precursor identity}
\label{sec:evap}

The gas is passive; volatility is limited by interfacial kinetics and by
conduction of latent heat through the film from the warm subphase
(one-sided model; \citealp{sultan2005}). Linearization of the
Hertz--Knudsen relation about saturation, with the pressure (Kelvin)
correction to the equilibrium vapour density (cf.\
\citealp{moosman1980}; \citealp[][their Eq.~(65)]{malachtari2025}),
gives the dimensionless flux
\begin{equation}
J=\frac{\Delta\,p+1}{K\,h+1}\,.
\tag{J}
\label{eq:J}
\end{equation}
The numerator is the unit superheat drive corrected by $\Delta\,p$: a
large negative film pressure depresses the equilibrium vapour density
and throttles evaporation. The denominator puts kinetic resistance
(the 1) in series with conductive resistance through the film ($Kh$).
Two structural consequences follow.

\textbf{(i) The precursor thickness is the $J=0$ isotherm.} In a flat
adsorbed film the curvature term in \eqref{eq:P} vanishes, so
$p=\Pi(h_p)=-A/h_p^3$, and $J=0$ requires $p=-1/\Delta$. Hence
\begin{equation}
h_p=(A\,\Delta)^{1/3}\,.
\tag{K}
\label{eq:K}
\end{equation}
The code's \code{hp = (A*Delta)**(1/3)} is a \emph{derived identity},
the direct analogue of \citet{malachtari2025} Eq.~(67) with a
one-term disjoining pressure and zero ambient humidity. The attractive
sign of $\Pi$ is forced: a repulsive (stabilizing) disjoining pressure
gives $p>0$ in the thin film, and the precursor evaporates without
bound. The precursor is an adsorbed film in thermodynamic equilibrium
with the vapour, in the tradition of Ajaev-type models. The equilibrium
is \emph{stable} despite $\Pi'(h_p)>0$ (the dewetting-unstable sign for
a passive film): a perturbation $\delta h>0$ relaxes $p$ towards zero
and unthrottles evaporation, restoring the film, and condensation
symmetrically restores $\delta h<0$. The evaporative restoring rate
dominates the van der Waals dewetting growth rate at the precursor
scale, the latter suppressed by $h_p^3$. This is a scaling statement;
we have not chased the prefactors.

\textbf{(ii) $A$, $\Delta$ and $h_p$ form a single parameter.} Any
sensitivity study that moves $h_p$ must move $A$ as $A\propto h_p^3$ (at
fixed $\Delta$). Violate \eqref{eq:K} and the precursor stays volatile
($J\neq0$ at $h=h_p$): a spurious mass source/sink parked exactly where
the deposit forms. The protocol: $h_p$ is moved through $A$ ($A\to8A$
for $h_p\times2$, $A/8$ for $h_p/2$), with the numerical smoothing
$h_{\mathrm{reg}}=h_p/10$ slaved and subdominant throughout.

\textbf{Known limitation, stated for the paper:} \eqref{eq:J} assumes
conduction-limited, superheat-driven volatility appropriate to a warm
subphase. Room-temperature decane is vapour-diffusion-limited; the
generalized diffusion-limited closure of \citet{sultan2005}, as
implemented by \citet{malachtari2024}, is the planned refinement.
The closure \eqref{eq:J} also differs structurally from that of
\citet{karapetsas2016}, whose flux carries the film thickness
directly in the resistance ($K+h$ rather than $Kh+1$); the validation of
Appendix~\ref{sec:benchmark} therefore benchmarks the present solver
\emph{under their closure} against their published dynamics. The closure
used here is a deliberate conduction-in-series model, not an
approximation of theirs.
The direction claim does not rest on the closure. Evaporation enters
the species problem only through $h$ (Remark~\ref{rem:nosource}): no
flux profile, however deformed, appears in \eqref{eq:Ti}, so any
closure that concentrates particles by thinning the film preserves the
species-symmetry of the operator and cannot touch the ordering. More precisely (\S\ref{sec:expansion}), the deposit composition is the
$O(D_i)$ term of an expansion in dimensionless diffusivity, and its sign
is fixed by $\operatorname{sign}(\partial_r h)$ at the front; every
closure producing a monotone inward-receding film shares it. For the
flat-topped lens the vapour flux is edge-peaked, so the film dries
edge-first and recession is monotone inward, the one condition the sign
argument requires. What a
different closure does change is the front kinematics $\dot r_f$, and
through it the spike widths of \eqref{eq:delta}: magnitude channel,
conceded. \citet{dambrosio2026} show that the CA-mode \emph{flow
direction itself} switches between uniform and diffusion-limited flux
profiles; the deposit \emph{shape} is therefore closure-dependent.
The \code{EVAP\_A} family checks directly that the
composition ordering is not.
The family deforms the flux \emph{multiplicatively},
$J\mapsto J\big(1+a\,w(r)\big)$ with $w$ a smooth profile supported over
the lens ($0\le r\lesssim1$) and members spanning edge-enhanced to
centre-enhanced flux, by design: a multiplicative
deformation preserves the $J=0$ root at $p=-1/\Delta$, so the precursor
remains inert under every member of the family. The invariance test
never confounds flux-profile deformation with a violated
identity~\eqref{eq:K}.

The test was run with $w$ centred on the lens interior at
$a\in\{0.5,1,2\}$ and with $w$ edge-peaked at $a=2$, against the $a=0$
baseline, all at $N_{\mathrm{el}}=1000$. In every member that reaches
dry-out the front recedes monotonically inward, so the precondition of
the sign argument holds; the edge-enhanced member does not dry within
the simulated window and is treated separately below. The composition
ordering never reverses in any member.

The centre-biased members and the baseline reach dry-out and carry the
quantitative claim; dry-out ranges from $t_{\mathrm{dry}}=62$ at the
strongest centre-bias to $189$ at baseline.
Separations here are
area-weighted mean-radius separations of the extreme species pair
\emph{in the three-species configuration}
($\Pe_{\mathrm{eff}}=9.7\times10^3$ against $3.6\times10^4$; the same
run as figure~\ref{fig:enrichment}), evaluated over $r\le1.05$ and read
at each member's own dry-out; the deposit is not frozen after dry-out
(\S\ref{sec:deposit}), so a common readout time would confound the
flux-profile effect with post-dry-out age. They are correspondingly
smaller than the five-species extreme-pair separation quoted in
\S\ref{sec:gate}, whose extreme pair spans a wider contrast, and the
two are not directly comparable. The separation falls monotonically
with centre-bias, $0.122$,
$0.082$, $0.068$, $0.034$ at $a=0,\,0.5,\,1,\,2$: compressed roughly
fourfold at the strongest centre-bias tested, still resolved (peak
positions certified to $0.002$, Appendix~\ref{sec:convergence}), and
never reversed. Flux bias is thus a signed magnitude knob; centre-bias
compresses the separation, and edge-bias, the opposite end of the same
monotone trend, sharpens it.

The edge-enhanced member does not reach dry-out. The multiplicative
enhancement is supported over the lens, so once the rim dries to the
precursor the enhancement no longer acts, and the thick residual cap
then evaporates slowly, throttled by its own thickness through the
$Kh+1$ resistance of \eqref{eq:J}. Its centreline height is
non-monotonic in time---$h(0)$ falls, rebounds, and falls again as
capillary refilling of the narrowing lens competes with
thickness-throttled evaporation---reaching $h(0)\approx0.47$ at $t=300$,
still far above the precursor. No dried deposit forms, so no
dried-deposit separation is defined for this member and none is quoted;
a separation read off the wet, oscillating lens is not a monotone bound
on any final value. The ordering is un-reversed throughout. The edge
member therefore confirms the direction claim qualitatively while lying
outside the quantitative magnitude comparison, which the centre series
carries on its own. The direction is fixed throughout, as the expansion
of \S\ref{sec:expansion} requires.

\subsection{Species transport: the conservative reduction}
\label{sec:species}

\subsubsection{Rapid vertical diffusion}
\label{sec:vertical}

With $\eps^2\Pe_i\ll1$ (\S\ref{sec:scaling}), write
$c_i=\phi_i(r,t)+\eps^2\Pe_i\,c_i^{(1)}(r,z,t)$; the $O(1)$ vertical
balance $\partial_z(D_i\partial_z c_i)=0$ with no-flux at both
interfaces forces $\phi_i$ to be independent of $z$
\citep{jensen1993,espin2014}. The concentration is vertically mixed;
all structure is radial.

\subsubsection{Depth integration: why evaporation never appears}
\label{sec:depthint}

Integrate the species equation across $s<z<\eta$ using the Leibniz
rule, the kinematic conditions, and the no flux
conditions~\eqref{eq:noflux}. Every boundary term cancels
\emph{identically}, including the one from the evaporative recession
of the top interface, because the particle flux condition there is
relative to the moving interface. What survives is
\begin{equation}
\partial_t\big(h\,\phi_i\big)
 +\frac{1}{r}\,\partial_r\!\Big[r\Big(
 \ubar\,h\,\phi_i-\frac{h}{\Pe_i}\,\partial_r\phi_i\Big)\Big]=0.
\tag{T$_i$}
\label{eq:Ti}
\end{equation}
The prognostic variable is the \textbf{areal load} $q_i=h\phi_i$; with
zero-flux boundary conditions, $\int q_i\,r\dd r$ is conserved
\emph{exactly} (in the FEM discretization, to machine precision)
because the weak form telescopes. The concentration $\phi_i=q_i/h$ rises
where $h$ thins. Evaporation concentrates particles through the
geometry, never through a source term.

\begin{remark}[the non-conservative trap]\label{rem:trap}
Evolving $\phi_i$ directly requires
$\partial_t\phi_i+\ubar\,\partial_r\phi_i=(EJ/h)\phi_i+\dots$: a source
term proportional to the evaporative flux, plus a diffusion operator
whose $h$-weighting must be inserted by hand. Mis-weight the diffusion
($1/(\Pe\,h)$ in place of $h/\Pe$) and a spurious flux $\propto\nabla h$
appears. That flux is enormous at the front, and in the present system
it \emph{reverses the predicted sorting direction}. We committed this
error during model development, diagnosed it, and record it here
as a warning: in front-dominated deposition problems, the conservative
form is a correctness requirement, not a style preference.
\end{remark}

\subsubsection{Structural theorem (mechanism attribution)}
\label{sec:theorem}

In \eqref{eq:Ti}, every species shares the \emph{identical} operator
except for the scalar $\Pe_i$: the same $\ubar$ (through $\mu(\ctot)$ and
$\Pdrv$, both functions of collective quantities) and the same geometry
$h$. Therefore \textbf{any compositional structure in the deposit is
attributable to diffusivity contrast and nothing else.} This is the
formal statement of the staged-activation principle: pinning
(\code{S\_PIN}), steric exclusion, trapping, jamming, Marangoni stress
and compliance are, by construction, perturbations to a baseline in
which $\Delta D$ is the \emph{only} symmetry-breaking between species.
Each acts through a collective field ($\ctot$, $h$, $\sigma$, $s$) that
every species sees identically, so each can move the magnitude of the
separation and none can move its sign.

\subsubsection{Mobility gate}
\label{sec:gate}

At extreme local loading the suspension jams; the code multiplies the
advective flux by
$g(\ctot)=\tfrac12\big(1-\tanh4(\ctot/c_{\mathrm{jam}}-1)\big)$, a
smooth arrest acting on the \emph{collective} concentration. The gate is
size-blind, so it is direction-neutral: it compresses contrasts; it
cannot create or reverse them. The gate multiplies the species
advective fluxes but \emph{not} the film mobility $h^3/3\mu$: a jammed
particle assembly arrests particle transport while the carrier fluid
keeps draining through it. Both operators remain size-blind.

The gate is \emph{active}, not inert: $\ctot$ at the front spike
crosses $c_{\mathrm{jam}}=50$ from $t\approx20$ onward and rises well
above it thereafter (the spike-body value in figure~\ref{fig:spike}(a)
is $O(10^2)$; the collective $\ctot$ peaks higher,
$O(10^3)$--$O(10^4)$, at the $h\to h_p$ shoulder, where
$\phi_i=q_i/h$ inherits the vanishing film thickness), so $g\to0$ at
the spike over most of the sweep. Its effect on the deposit is
nonetheless small, and size-blind. Against a gate-off control
($c_{\mathrm{jam}}\to\infty$, otherwise identical, five species,
$N_{\mathrm{el}}=1000$), the per-species mean deposit radii shift by at
most $6.8\times10^{-3}$ (gate on, $\langle r\rangle_i=0.392$, $0.473$,
$0.519$, $0.540$, $0.548$; gate off, $0.394$, $0.479$, $0.524$,
$0.542$, $0.548$) and the extreme-pair separation by about $1\%$
($0.1559$ against $0.1544$), with the species ordering and the sorting
direction identical in the two runs. Two conclusions follow. Deposit
freezing is dominated by the $h\to h_p$ collapse of the film mobility
$h^3/3\mu$, not by jamming: removing the gate entirely leaves the
deposit essentially where it was. And the gate, a function of the
collective $\ctot$ alone, cannot create or reverse size sorting --- the
control demonstrates directly what \S\ref{sec:theorem} guarantees
structurally.

A related numerical device: the diffusion floor
$D_i=(1+\delta_{\mathrm{rel}})/\Pe_i+D_{\mathrm{abs}}$ compresses the
diffusivity contrast asymmetrically (the floor is a larger fraction of
$D$ for the least diffusive species). All quantitative comparisons in
\S\ref{sec:front} therefore use the \emph{effective} P\'eclet number
$\Pe_{i,\mathrm{eff}}=1/D_i$, reported with every run, rather than the
nominal $\Pe_i$. The floor biases \emph{against} the segregation
signal. It cannot manufacture the effect it is used to measure.
$D_{\mathrm{abs}}$ is set \emph{per run} as a fixed fraction of the
smallest physical diffusivity in that run,
$D_{\mathrm{abs}}=0.1/\max_i\Pe_i$, so $\Pe_{\mathrm{eff}}$ is a
per-run quantity and is not comparable across configurations at fixed
nominal $\Pe$: the nominal $\Pe=10^4$ species carries
$\Pe_{\mathrm{eff}}=9.7\times10^3$ in the three-species runs
(figures~\ref{fig:enrichment}, \ref{fig:dynamics}) and
$9.9\times10^3$ in the five-species runs
(figure~\ref{fig:scaling}). Every figure quotes the
$\Pe_{\mathrm{eff}}$ of its own run.

\emph{Not adopted, and why:} the concentration-dependent
hindered/osmotic diffusivity $D(\Phi)$ of \citet{malachtari2025} (their
Eq.~(28), non-monotonic, divergent at packing) is size-symmetric in
structure but would soften spikes exactly where they live. It belongs
in the magnitude sensitivity discussion, not the baseline operator.
Adopting it would blur the attribution theorem of \S\ref{sec:theorem}
for no gain in the direction claim.

\subsubsection{The composition expansion}
\label{sec:expansion}

The theorem of \S\ref{sec:theorem} attributes any deposit composition
to diffusivity contrast. It is worth making that constructive, because
the dimensionless diffusivity $D_i=1/\Pe_{i,\mathrm{eff}}\ll1$
(equation~\ref{eq:regime}) is a small parameter: expand \eqref{eq:Ti}
in it.

\emph{Base state ($D_i\to0$).} With diffusion off, \eqref{eq:Ti} is pure
advection, $\partial_t q_i+\tfrac1r\partial_r(r\,\ubar\,q_i)=0$, and
$\ubar$ is collective (through $\mu(\ctot)$ and $\Pdrv$) and therefore
common to all species. The loading \eqref{eq:icload} is
$q_i(r,0)=f_i\,\Theta(r)$ with the profile $\Theta$ shared: a common
linear operator on proportional data separates,
$q_i^{(0)}(r,t)=f_i\,Q^{(0)}(r,t)$, one size-blind field. The
composition $\phi_i^{(0)}/f_i=Q^{(0)}/h$ is then flat in species: no
sorting at zeroth order. (This promotes the compositional symmetry of
the initial condition, \S\ref{sec:icbc}, to all times at leading order.)

\emph{First order.} The only $i$-dependent term in \eqref{eq:Ti} is the
diffusive flux $-D_i\,h\,\partial_r\phi_i$. Evaluated on the base state,
$\partial_r\phi_i^{(0)}=f_i\,\partial_r(Q^{(0)}/h)$, so the flux is
$D_i f_i\,\mathcal{F}(r,t)$ with $\mathcal{F}\equiv-h\,\partial_r(Q^{(0)}/h)$
\emph{common} to all species. Hence $q_i^{(1)}=f_i\,Q^{(1)}$ with
$Q^{(1)}$ common, and
\begin{equation}
q_i \;\approx\; f_i\big[\,Q^{(0)}+D_i\,Q^{(1)}\,\big]\;+\;O(D_i^2).
\label{eq:qexpand}
\end{equation}
Every species shares one shape at each order and differs only through
the scalar $D_i$. This is \S\ref{sec:theorem} made explicit: the deposit
composition \emph{is} the $O(D_i)$ term, and it inherits its sign from
$\mathcal{F}$.

\emph{Sign, and why the closure cannot set it.} Decompose the
size-dependent flux exactly:
\begin{equation}
-D_i\,h\,\partial_r\!\big(q_i/h\big)
\;=\;\underbrace{-D_i\,\partial_r q_i}_{\text{Fickian}}
\;+\;\underbrace{D_i\,\phi_i\,\partial_r h}_{\text{shoulder rectifier}}.
\label{eq:fluxdecomp}
\end{equation}
The Fickian term drifts the (area-weighted) mean radius \emph{outward} in
the cylindrical first moment: geometric spreading, identical across
species up to the scalar $D_i$. The rectifier carries $\partial_r h$,
which is one-signed at a monotone receding front (the film thins
outward, $\partial_r h<0$), and it is weighted by $\phi_i$, largest at
the front spike.
The rectifier dominates. Resolving the two terms of
\eqref{eq:fluxdecomp} as separate contributions to the deposit moment
rate $\dd\langle r\rangle_i/\dd t$ over $r\le1.05$ (five-species
production run, $N_{\mathrm{el}}=1000$; diagnostic
\code{flux\_decomp.py}), the ratio
$|\text{rectifier}|/|\text{Fickian}|$ is independent of $D_i$, which
cancels: it is a property of the profile geometry alone. It exceeds
unity for every species at every time sampled across the recession
sweep ($t=60$--$180$), ranging from $2.2$ to $10.5$ and peaking near
mid-recession.
The net $O(D_i)$ transport is therefore inward, and larger
for the more diffusive species:
$D_{\mathrm{small}}>D_{\mathrm{large}}\Rightarrow$ small-rich centre,
large-rich annulus. The direction is fixed by $\operatorname{sign}
(\partial_r h)$ at the front and by nothing in the evaporation closure:
the flux profile and the flow enter \eqref{eq:qexpand} only through the
size-blind magnitudes $\ubar$ and $\dot r_f$ that shape $Q^{(0)},Q^{(1)}$
(the magnitude channel). Any closure producing a monotone inward-receding
film gives the same sign.

\emph{Outer solution.} The expansion \eqref{eq:qexpand} is regular in the
bulk but \emph{non-uniform} at the front: inside the layer
$\delta_i\sim D_i/|\ubar-\dot r_f|$ (\S\ref{sec:front}) the correction
$D_i\,\partial_r\phi_i\sim\phi_{\mathrm{spike}}\,|\ubar-\dot r_f|=O(1)$,
not $O(D_i)$. So \eqref{eq:qexpand} is the outer solution; the deposit
composition, an integrated (abandoned) quantity, is genuinely $O(D_i)$,
and \S\ref{sec:front} resolves the inner spike and matches to it.

\emph{Resolution law.} From \eqref{eq:qexpand} the mean deposit radius is
\begin{equation}
\langle r\rangle_i\;\approx\;\langle r\rangle^{(0)}+D_i\,\Delta^{(1)},
\qquad
\Delta^{(1)}=\frac{\displaystyle\int r\,Q^{(1)}\,r\dd r}
{\displaystyle\int Q^{(0)}\,r\dd r},
\label{eq:meanr}
\end{equation}
a single common slope $\Delta^{(1)}$ (negative: larger $D$, smaller
$\langle r\rangle$). The adjacent-pair separation follows immediately,
$|\langle r\rangle_i-\langle r\rangle_j|\approx|\Delta^{(1)}|\,
\Delta(1/\Pe_{\mathrm{eff}})_{ij}$: linear in the
inverse-P\'eclet contrast, through the origin, with the measured slope
of \S\ref{sec:bandsep} identified as $|\Delta^{(1)}|$. That is
Prediction~3 of
\S\ref{sec:front}, now \emph{derived}, and the neglected $O(D_i^2)$
term bounds departures from linearity at large contrast
(\S\ref{sec:bandsep}).

\subsection{Front boundary layer and the chromatograph scalings}
\label{sec:front}

Bulk radial diffusion is $O(1/\Pe_i)$: negligible. The mechanism lives
in a self-selected boundary layer at the receding front. Let $r_f(t)$
be the front position (the $h\to h_p$ shoulder) with $\dot r_f<0$, and
pass to the front frame, $\xi=r-r_f(t)$. In \eqref{eq:Ti} the advective
flux relative to the front is $q_i(\ubar-\dot r_f)$; a quasi-steady
spike forms where this is balanced by diffusion:
\begin{equation}
\big(\ubar-\dot r_f\big)\,h\,\phi_i
 \;\sim\;\frac{h}{\Pe_i}\,\partial_\xi\phi_i
\quad\Longrightarrow\quad
\delta_i\sim\frac{1}{\Pe_i\,|\ubar-\dot r_f|}
\label{eq:delta}
\end{equation}
The balance \eqref{eq:delta} is written for the \emph{ungated}
advective flux. Since the gate is saturated closed at the peak
(\S\ref{sec:gate}), the width is set on the spike flanks, where $g$ is
unsaturated; the gate-off control of \S\ref{sec:gate}, which reproduces
the deposit to $7\times10^{-3}$ in $\langle r\rangle_i$, certifies that
this distinction is immaterial to the results reported.
This is an exponential spike of width $\delta_i\propto D_i$. The more
diffusive (smaller) species piles up wide and shallow; the less
diffusive (larger) species is locked in narrow and tall. The deposit at
radius $r$ is the areal load abandoned as the front sweeps past,
$q_i^{\mathrm{dep}}(r)\approx q_i\big(r,t_f(r)\big)$ with $t_f$ the
passage time. The conversion of leakage into radial ordering rests on
\emph{which fluid survives the front's passage}. The wide spike's
diffusive tail extends inward, into liquid that outlives the front and
rejoins the interior reservoir. The narrow spike's mass is confined to
the annulus being abandoned. So the more diffusive species keeps
leaking backward out of the moving spike. The swept annulus ends up
enriched in the large species; the last liquid at the centre, in the
small. Direction:
$D_{\mathrm{small}}>D_{\mathrm{large}}$ (Stokes--Einstein)
$\Rightarrow$ small-rich centre, large-rich annulus. Every ingredient
upstream of this argument (flux profile, rheology, gate, subphase)
enters only through $|\ubar-\dot r_f|$ and $h$, both collective,
size-blind quantities. That is the structural reason the direction
claim is robust while any magnitude claim is model-dependent.

\textbf{Pre-registered $N$-species predictions}, fixed before the
production runs and tested against \code{front\_tracks\_Nsize} output
rather than fitted after the fact:
\begin{enumerate}
\item Spike widths narrow monotonically with $\Pe_{i,\mathrm{eff}}$;
the absolute widths are floored by the film-front shoulder width, which
the concentration $\phi_i=q_i/h$ inherits through $1/h$, so the
diffusive sub-layer of \eqref{eq:delta} is not separately resolved in
the instantaneous profile (\S\ref{sec:frontal}). Amplitudes are
anti-ordered, subject to the capture/leakage crossover (early times are
capture-dominated and can transiently invert the amplitude ordering, as
observed at $t=2$; the late-time ordering has small above large).
\item Deposit bands radially ordered monotonically in
$\Pe_{\mathrm{eff}}$.
\item Band separation for adjacent species
$\Delta r_{ij}\propto\Delta(1/\Pe_{\mathrm{eff}})_{ij}$ at fixed sweep
history, degrading when $\Delta(1/\Pe_{\mathrm{eff}})$ approaches either
the diffusion floor or the front-width smearing scale
$\max(\delta_i,\delta_j)$: the \textbf{resolution limit} of the
contact-line chromatograph. If clean, this is the paper's central
figure. The linear law and its saturation are derived as the
outer-moment relation \eqref{eq:meanr} of \S\ref{sec:expansion}.
\end{enumerate}

\emph{Relation to prior asymptotics:} this layer is the moving-front,
multispecies generalization of the contact-line diffusive boundary layer
analysed for pinned droplets by \citet{moore2021}. In that problem the
layer supplies a correction to the coffee ring; here it supplies the
entire mechanism.

\subsection{Initial and boundary conditions}
\label{sec:icbc}

\begin{equation}
h(r,0)=h_p+\max\!\big(1-r^2-h_p,\,0\big),
\qquad
s(r,0)=-\Rrho\,\big(h(r,0)-h_p\big),
\label{eq:ich}
\end{equation}
the latter being isostatic flotation of the lens measured relative to
the equilibrium precursor (Appendix~\ref{sec:winkler}), exactly
compatible with the far-field condition $s(L_x)=0$. The slope
discontinuity of the initial profile at $r=1$ is smoothed by
capillarity within $O(\Ct)$ time units, well before front formation.
\begin{equation}
q_i(r,0)=h(r,0)\;C_{\mathrm{tot}}\,f_i\;
\tfrac12\big(1+\tanh k_L(R_{\mathrm{load}}-r)\big),
\qquad
\textstyle\sum_i f_i=1,
\label{eq:icload}
\end{equation}
with $R_{\mathrm{load}}=0.92$ and steepness $k_L=10$ (\code{LOAD\_K}).
This profile carries a loading tail: at $r=1$ it stands at $0.17$ of
the centre value, so a small fraction of the particle mass is seeded
outboard of the lens at $t=0$. A steeper control at $k_L=30$, whose
tail is $O(10^{-5})$ by $r=1.1$, changes the dried rim mass from
$2.0\%$ to $1.9\%$ of the total: the loading tail deposits no
mass-carrying rim, and the faint outer rim of
figure~\ref{fig:deposit5} is a composition feature of the precursor
tail, not a loading artefact.
The fractions $f_i=\code{FRAC}$ hold the total load fixed across $N$ so
that $\mu(\ctot)$ and $g(\ctot)$ see identical collective fields in any
$N$-to-$N$ comparison (\S\ref{sec:theorem}). Identical loading profiles
across species make the initial condition compositionally symmetric:
any composition difference in the dried centre is transport. Boundary
conditions: regularity at $r=0$; at $r=L_x=4$: zero flux for $h$ and
$q_i$, $s=0$ (far-field EG reservoir), precursor equilibrium via
\eqref{eq:K}.

\begin{figure}[t]
  \centering
  \safeincludegraphics[width=0.75\linewidth]{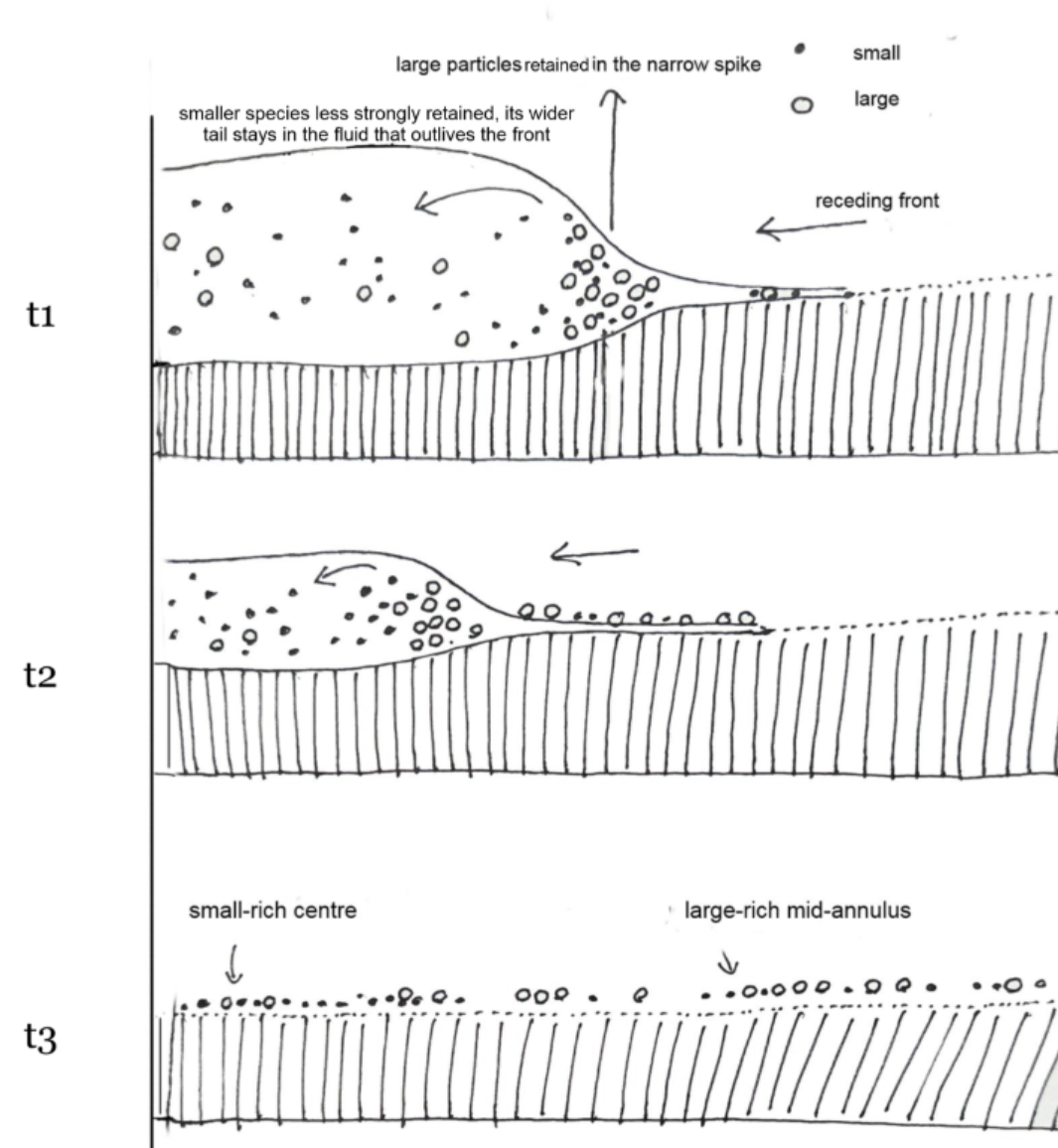}
  \caption{Proposed frontal-fractionation mechanism (schematic). As the
    front recedes toward the axis ($t_1\!\to\!t_3$), both species crowd
    into a concentration spike at the moving front. The larger species
    is held tightly in the narrow spike and is laid down along the sweep
    path. The smaller, more diffusive species is held loosely; its wider
    spike leaves more mass in the fluid that outlives the front. The
    time-integrated result: small-rich centre, large-rich mid-annulus.}
  \label{fig:schematic}
\end{figure}

\section{Results}
\label{sec:results}

\subsection{Frontal fractionation at the receding contact line}
\label{sec:frontal}

\begin{figure}[tp]
  \centering
  \safeincludegraphics[width=\textwidth]{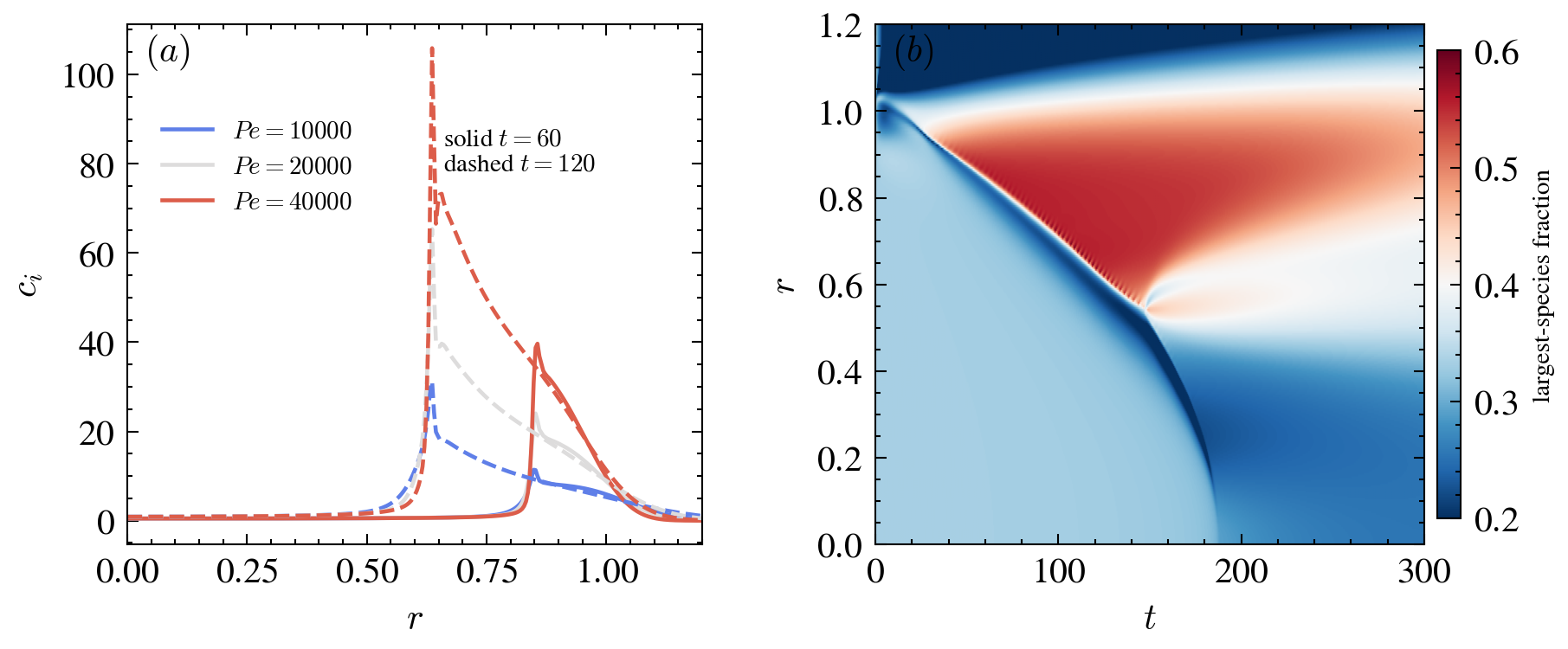}
  \caption{Frontal fractionation at the receding contact line.
    (\textit{a}) Particle concentration $c_i(r)$ near the front for
    three P\'eclet numbers ($\Pe = 10^4,\,2\times10^4,\,4\times10^4$),
    at $t=60$ (solid) and $t=120$ (dashed). All species pile into a
    single spike co-located with the receding front (agreement to
    $0.003$ in position). The spike sharpens and grows as it sweeps
    inward, and its width narrows monotonically with $\Pe$
    ($0.17\to0.10$ at $t=60$); the widths are of order the film-front
    shoulder that the concentration inherits through
    $\phi_i=q_i/h$ (see \S\ref{sec:frontal}). Widths are quoted at
    $t=60$, while all three spikes are isolated; by $t=120$ the
    largest-$\Pe$ spike has begun merging with its own deposit and its
    width is no longer a clean measure.
    (\textit{b}) Space--time map of the largest-species fraction. The
    front-localized couplet sweeps inward as the lens dries; the final
    deposit is the time-integrated record of the sweep.}
  \label{fig:spike}
\end{figure}

All sorting in this system is written at the front. As the film dries, the
contact line recedes toward the axis, and the depth-integrated species balance
(T$_i$) develops a concentration spike co-located with the receding shoulder.
Every species piles up where the film thins to the precursor.
Figure~\ref{fig:spike}(a) shows this spike for three P\'eclet numbers at two
times. The structure is common to all species: one spike, at the front
(spike location and the $h$-front agree to $0.003$ at both $t=60$ and
$t=120$), sharpening and growing as it sweeps inward.
Its width is not common: the spikes narrow monotonically with $\Pe$,
from $\delta\approx0.17$ to $0.10$ at $t=60$, quoted while all three
spikes are still isolated. Those widths are of order the film-front
shoulder width ($0.155$, 10--90 measure) and not of order the diffusive
sub-layer, because the concentration inherits the shoulder through
$\phi_i=q_i/h$: the spike is hosted by the shoulder and cannot be much
narrower than it, whatever \eqref{eq:delta} would give in isolation. The
quasi-steady balance of front-relative advection against diffusion,
\eqref{eq:delta}, predicts a sub-layer of width
$\delta_i\sim1/(\Pe_i\lvert\ubar-\dot r_f\rvert)$, of order
$0.01$--$0.04$ at the measured front-frame speed; that layer lies
inside the shoulder and is not separately resolved in the instantaneous
profile. The monotonic narrowing across the three P\'eclet numbers, at
a fixed shoulder, is its residual signature. The
quantitative content of \eqref{eq:delta} is read out downstream, in the
deposit band separations of \S\ref{sec:bandsep}, which integrate the
differential retention over the full sweep.

This width difference is the whole mechanism, because it decides which species
survives the front's passage. The wide spike's diffusive tail extends inward,
into liquid that outlives the front and rejoins the interior reservoir. The
narrow spike's mass is confined to the annulus being abandoned. So the more
diffusive species keeps leaking backward out of the moving spike, the swept
annulus ends up enriched in the large species, and the last liquid at the
centre ends up enriched in the small. The deposit at radius $r$ is the areal
load abandoned as the front passes, $q_i^{\mathrm{dep}}(r) \approx q_i\!\left(r,
t_f(r)\right)$. Figure~\ref{fig:spike}(b) reads this as a space--time record.
The largest-species fraction, plotted over $(r, t)$, shows the front-localized
couplet sweeping radially inward as the lens dries, so the final deposit is the
time-integrated trace of that sweep.

The direction of the ordering is fixed, and the argument for it is structural.
Every ingredient upstream of the spike enters \eqref{eq:delta} only through
$\lvert\ubar - \dot{r}_f\rvert$ and $h$. The flux profile, the rheology
$\mu(\ctot)$, the mobility gate $g(\ctot)$, and the subphase coupling are all
collective and size-blind. Size enters the transport problem through the
Stokes--Einstein diffusivity and nowhere else. By the symmetry argument of
\S\ref{sec:theorem}, any compositional structure in the dried film must
therefore come from the diffusivity contrast alone, and its direction is set by
the sign of that contrast: $D_{\mathrm{small}} > D_{\mathrm{large}}$ gives a
small-rich centre and a large-rich annulus. This is the license under which the
rest of \S\ref{sec:results} reads what it observes. Direction is a theorem.
Magnitude is a measurement.

\subsection{The dried deposit}
\label{sec:deposit}

\begin{figure}[tp]
  \centering
  \safeincludegraphics[width=\textwidth]{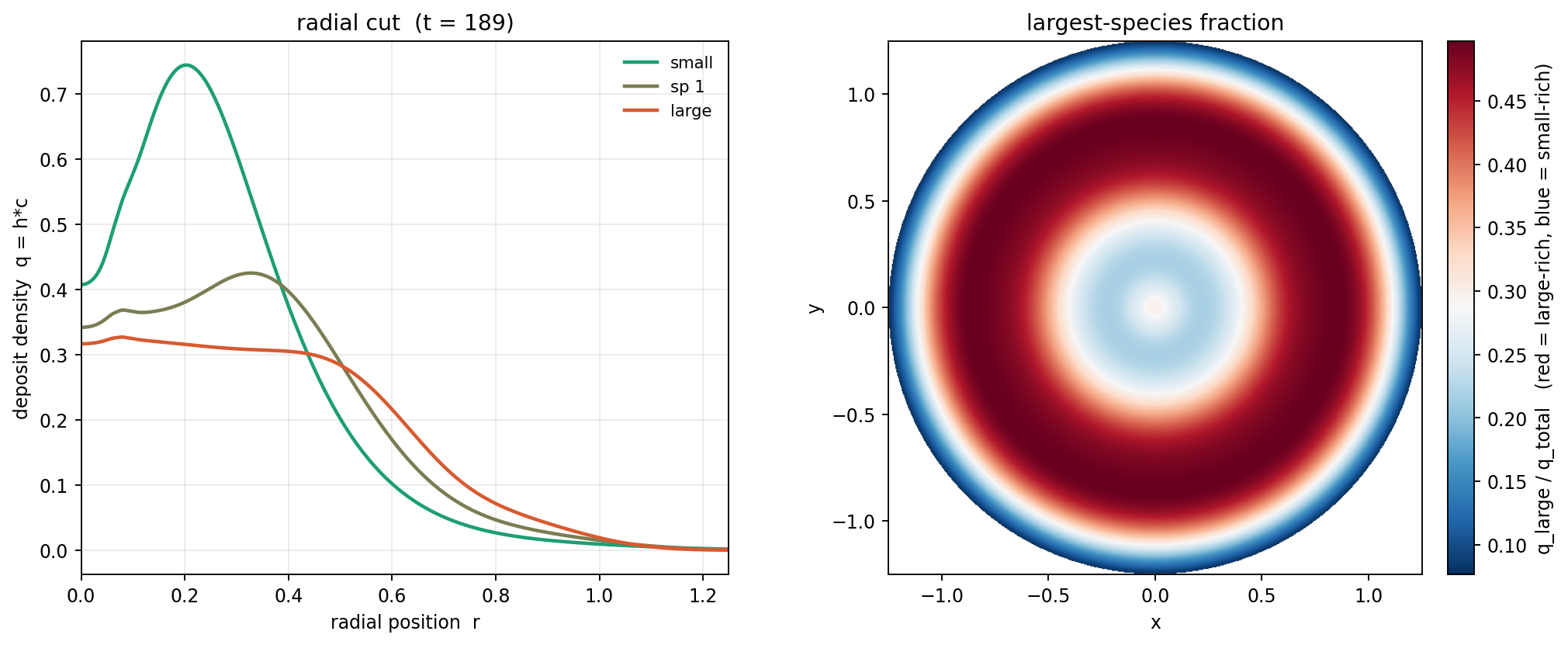}
  \caption{Final dried deposit, three species ($N_{\mathrm{el}}=1000$),
    read at dry-out, $t_{\mathrm{dry}}=189$: dry-out arrests advection
    (the film mobility $h^3/3\mu$ collapses as $h\to h_p$) but not
    diffusion, and the deposit smears slowly thereafter
    (\S\ref{sec:deposit}). \textit{Left:} deposit density
    $q_i(r)=h\,c_i$ per species. The smallest species peaks at the
    centre and is lowest at the edge; the largest carries the
    mid-annulus; the intermediate peaks between them. \textit{Right:}
    the same deposit as a top-down disk coloured by largest-species
    fraction: small-rich centre, large-rich mid-annulus. All species
    decay to zero by $r\!\approx\!1$. The deposit is two-zone
    \emph{in mass}, with no persistent precursor rim.}
  \label{fig:deposit3}
\end{figure}

Figure~\ref{fig:deposit3} shows the deposit this sweep writes, for a
three-species run ($N_{\mathrm{el}} = 1000$). The per-species radial cut
(figure~\ref{fig:deposit3}, left) resolves the predicted ordering. The smallest
species peaks at the centre and is lowest at the edge; the largest carries a
mid-radius annulus; the intermediate, between them. Rendered as
a top-down disk coloured by largest-species fraction
(figure~\ref{fig:deposit3}, right), the deposit is two-zone: a small-rich centre
and a large-rich mid-annulus, with every species decaying to zero by $r \approx
1$ and no persistent precursor rim carrying mass. This is the direction claimed
in \S\ref{sec:frontal}. It is also the reverse of the ordering established for
pinned bidisperse deposits on solids, where the small species is driven to the
contact line and the large toward the interior. Here the small species is
retained at the centre and the large is stranded in the annulus. Nothing about
the pinned-substrate result was assumed. The reversal is an output of the
substrate-free geometry.

All deposit quantities in this paper are read at dry-out,
$t_{\mathrm{dry}}=189$, defined as the time at which the receding
front reaches the axis and the film has thinned to the precursor over
the whole lens. The readout time matters because dry-out arrests
advection, not diffusion: the film mobility $h^3/3\mu$ collapses as
$h\to h_p$, but the species diffusion survives, and at uniform
$h=h_p$ it reduces to plain radial diffusion at
$1/\Pe_{\mathrm{eff}}$. The deposit therefore smears slowly after
dry-out, fastest for the most diffusive species: the Fickian term of
\eqref{eq:fluxdecomp} outliving the rectifier. Between $t=189$ and
$t=300$ the mean radius of the most diffusive species drifts from
$0.391$ to $0.427$ while the least diffusive is locked at $0.548$,
and the fitted resolution slope of figure~\ref{fig:scaling} decays
from $1.66\times10^{3}$ to $1.24\times10^{3}$. This is a limitation
of the model, which continues to treat the precursor-thick film as
liquid: a real dried deposit does not diffuse. The remedy is a
solidification gate on the species diffusivity,
$D_i\mapsto D_i\,\Lambda(h)$ with $\Lambda$ a smooth switch vanishing
as $h\to h_p$, structurally identical to the size-blind mobility gate
of \S\ref{sec:gate} and therefore direction-neutral by the same
argument. We have not adopted it here, because reading every deposit
quantity at the common dry-out time $t_{\mathrm{dry}}=189$ makes the
choice immaterial to the results reported.

\begin{figure}[tp]
  \centering
  \safeincludegraphics[width=\textwidth]{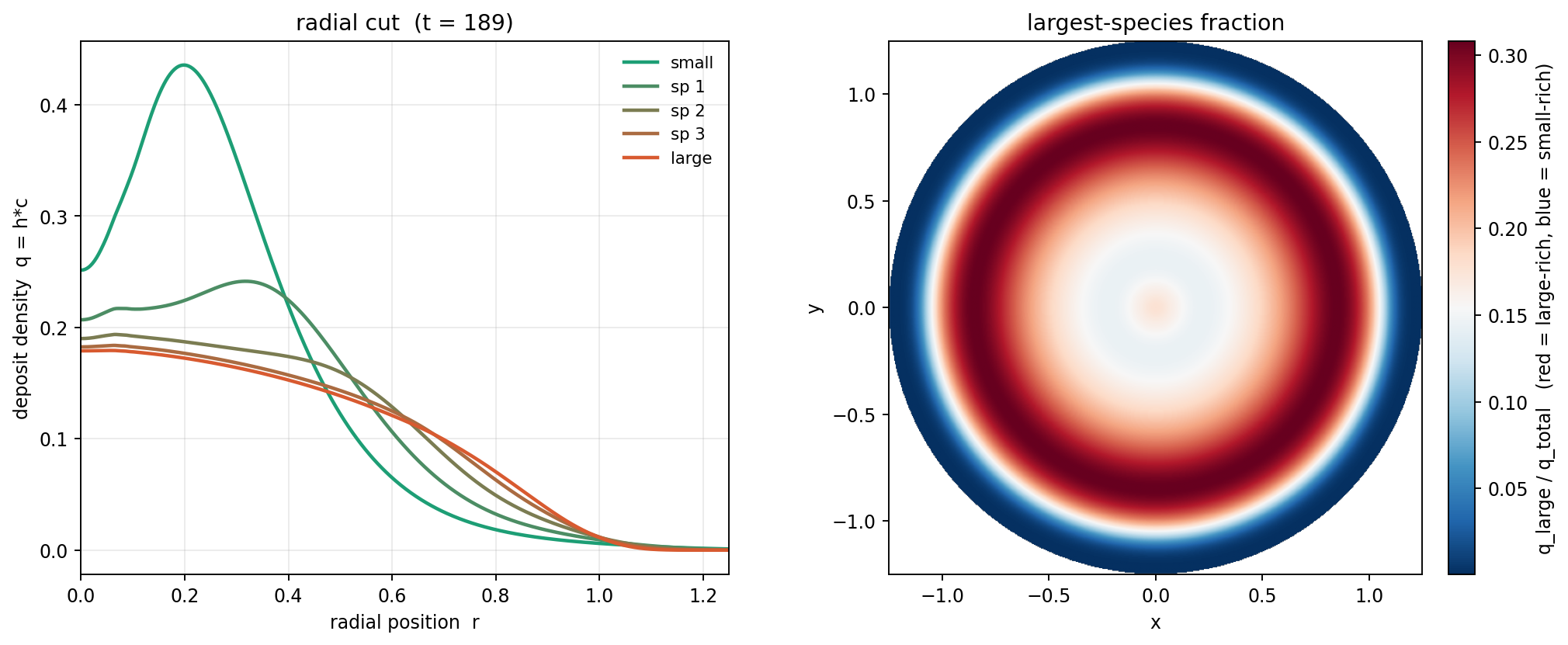}
  \caption{Final dried deposit, five species ($N_{\mathrm{el}}=1000$;
    $\Pe=10^4$ to $1.6\times10^5$; read at dry-out,
    $t_{\mathrm{dry}}=189$, \S\ref{sec:deposit}).
    \textit{Left:} the ordering at the centre is monotonic in size:
    the smallest sits highest, and each larger species sits below the
    last. The four largest species are closely spaced because $1/\Pe$
    compresses at large $\Pe$; the sorting does not. \textit{Right:}
    the pale centre and dark annulus are the mass-carrying zones. The
    blue outer rim is a composition feature of the precursor tail,
    where every species has decayed to $O(10^{-3})$ of its central
    value: visible in the fraction, absent in the mass. Prediction~2
    of \S\ref{sec:front}, tested at $N=3$ and $N=5$. The sorting is a
    property of the mechanism, not of the number of species used to
    sample it.}
  \label{fig:deposit5}
\end{figure}

Figure~\ref{fig:deposit5} repeats the experiment with five species
($N_{\mathrm{el}} = 1000$,
$\Pe = 10^4$ to $1.6\times10^5$) and shows that the three-species
picture is not an artefact of coarse sampling. The per-species deposit
densities are ordered monotonically in size at the centre: the smallest species
sits highest, and each larger species sits below the last. The four largest
species sit closely spaced, and the reason is arithmetic: $1/\Pe$
compresses at large $\Pe$. The sorting does not.
The mechanism keeps separating species whose diffusivity contrast is
small; the bands just crowd closer.
On the disk (figure~\ref{fig:deposit5}, right), the pale centre and
dark annulus are the mass-carrying zones. The blue outer rim is a composition
feature of the precursor tail, where every species has decayed to $O(10^{-3})$
of its central value, so it appears in the fraction but not in the mass
(and the loading control of \S\ref{sec:icbc} confirms the initial
tail contributes no rim mass).
Figure~\ref{fig:deposit5} is Prediction~2 of \S\ref{sec:front}, that
deposit bands are radially ordered monotonically in $\Pe_{\mathrm{eff}}$,
tested at $N = 5$ rather than $N = 3$ and confirmed.

\begin{figure}[t]
  \centering
  \safeincludegraphics[width=0.55\linewidth]{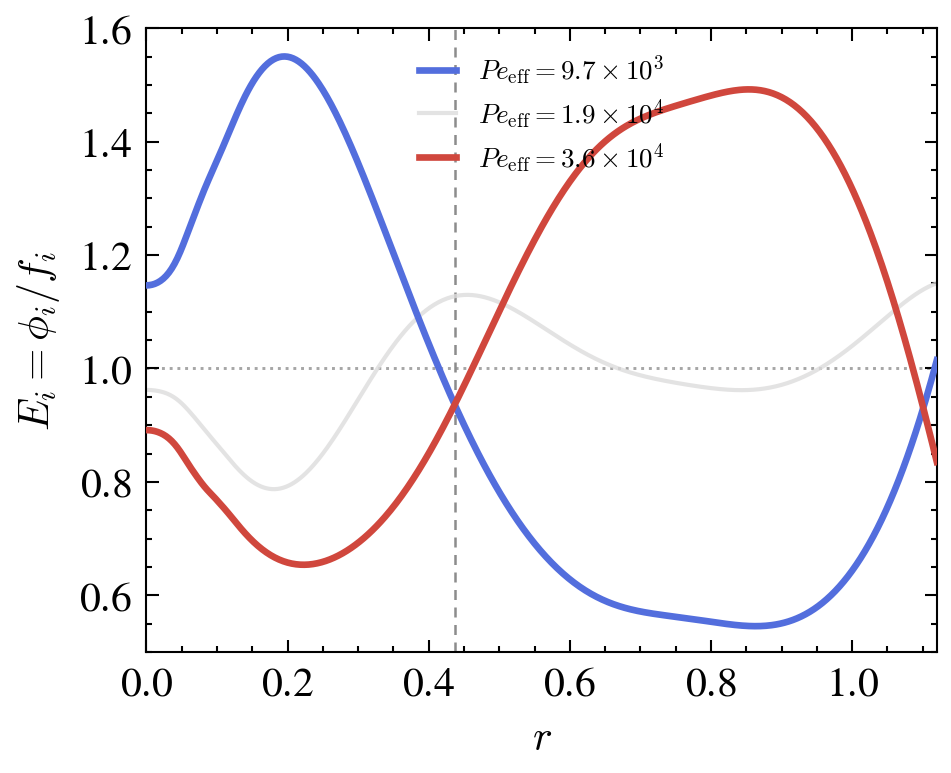}
  \caption{Enrichment $E_i(r)=\phi_i/f_i$ across the final deposit
    (three-species run; read at dry-out, $t_{\mathrm{dry}}=189$,
    \S\ref{sec:deposit}). $E_i>1$ marks over-representation relative
    to the loading fraction $f_i$. The smallest species
    ($\Pe_{\mathrm{eff}}=9.7\times10^3$) is enriched at the centre and
    depleted in the annulus; the largest
    ($\Pe_{\mathrm{eff}}=3.6\times10^4$) is depleted at the centre and
    enriched in the mid-annulus, peaking near $r\approx0.9$. The curves
    cross $E=1$ together near $r\approx0.44$: the composition-neutral
    radius separating small-rich centre from large-rich annulus. The
    direction claim of \S\ref{sec:front}, read directly off the
    deposit. $\Pe_{\mathrm{eff}}$ is quoted for this run; the
    diffusion floor is set per run (\S\ref{sec:gate}).}
  \label{fig:enrichment}
\end{figure}

Figure~\ref{fig:enrichment} reads the same three-species deposit as an
enrichment $E_i(r)=\phi_i/f_i$, the local over-representation of each
species relative to its loading fraction. The smallest species is
enriched at the centre and depleted in the annulus; the largest is
depleted at the centre and enriched in the mid-annulus, peaking near
$r\approx0.9$. The curves cross $E=1$ together near $r\approx0.44$: a
composition-neutral radius separating the small-rich centre from the
large-rich annulus. This is the direction claim of \S\ref{sec:front},
read directly off the deposit.

\subsection{Recession dynamics and band separation}
\label{sec:bandsep}

\begin{figure}[t]
  \centering
  \safeincludegraphics[width=0.6\linewidth]{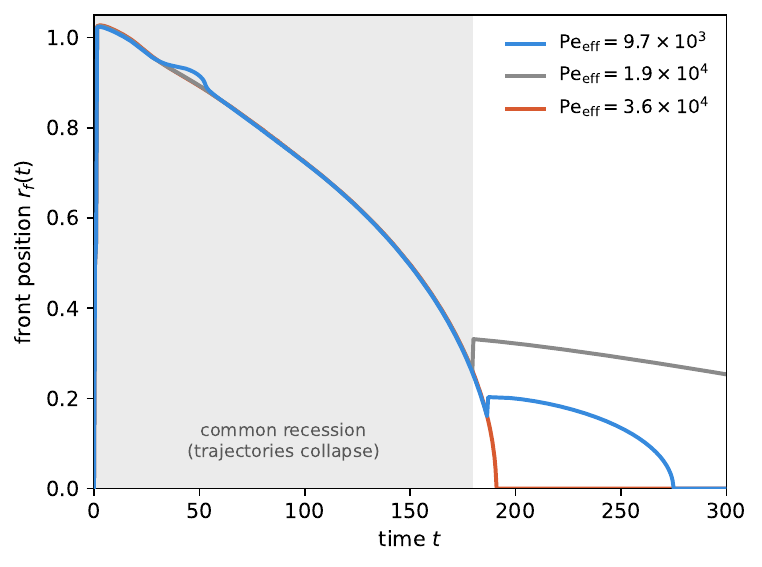}
  \caption{Front kinematics up to dry-out ($t_{\mathrm{dry}}\approx189$).
    The black curve is the film front (the $h\to h_p$ shoulder), a
    single collective object common to all species
    (\S\ref{sec:theorem}). Coloured curves track the per-species
    concentration fronts for three P\'eclet numbers. Through the
    loading phase ($t\lesssim150$) all species co-recede with the film
    front. In the final collapse the species fronts peel off it: the
    smallest runs to the axis, the largest is stranded in the
    mid-annulus.
    The peel-off is the endgame of the retention asymmetry, not its
    cause: the film front itself is a single object, and the species
    fronts separate from it only as each deposit freezes.
    Film recession is species-independent;
    the deposit ordering is written by differential retention at the
    common front, not by differential front motion.
    $\Pe_{\mathrm{eff}}$ values are those of the three-species run
    (\S\ref{sec:gate}).}
  \label{fig:dynamics}
\end{figure}

The mechanism of \S\ref{sec:frontal} attributes sorting to differential
retention within the front, not to the species moving the front differently.
Figure~\ref{fig:dynamics} tests that premise directly. Tracking the front
position $r_f(t)$ for each species across three P\'eclet numbers, the
trajectories collapse onto a single curve. Recession is species-independent to
within the tracking resolution. The front is one object. Small and large
species ride the same receding shoulder; at any given radius they are
laid down at the same instant. What differs is only how strongly the
shoulder retains each as it passes.
This rules out the obvious alternative to
the boundary-layer account, that the species sort because they recede at
different rates and so deposit at different radii. They do not recede at
different rates. The sorting of figures~\ref{fig:deposit3}--\ref{fig:enrichment}
is entirely a retention effect within a common front, exactly as
\eqref{eq:delta} requires. Every ingredient upstream of the spike enters
through the size-blind quantities $\lvert\ubar - \dot{r}_f\rvert$ and
$h$; the front dynamics cannot carry size information on their own.
The same moment-rate decomposition makes the endgame visible directly.
At $t=180$, just before dry-out, the size-dependent contribution to
$\dd\langle r\rangle_i/\dd t$ is $-2.1\times10^{-5}$ for the largest
species against $-1.6\times10^{-3}$ for the smallest: the large
species' inward drift has collapsed toward zero as its deposit freezes
into the annulus, while the small species is still being drawn toward
the axis some eighty times faster.
The deposit ordering is the time-integrated record of
exactly that asymmetry.

\begin{figure}[tp]
  \centering
  \safeincludegraphics[width=\textwidth]{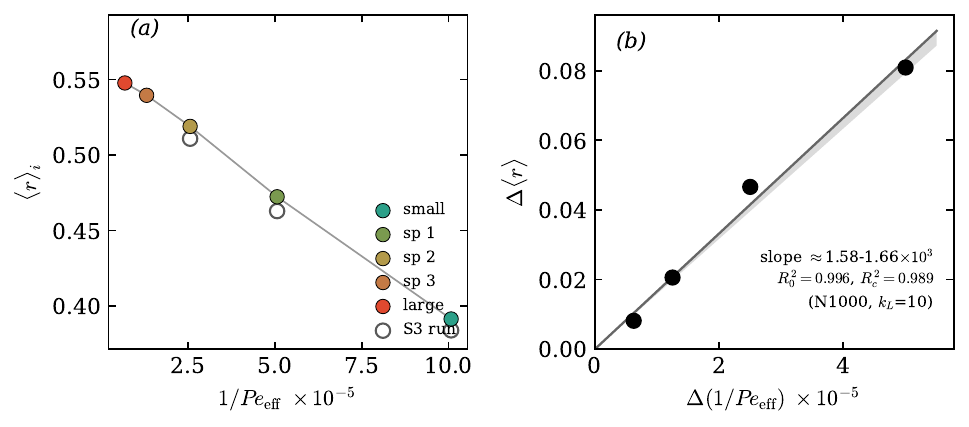}
  \caption{Band separation scales with diffusivity contrast (five
    species, $N_{\mathrm{el}}=1000$; deposit read at dry-out,
    $t_{\mathrm{dry}}=189$, \S\ref{sec:deposit}).
    (\textit{a}) Mean deposit radius $\langle r\rangle_i$ against
    $1/\Pe_{\mathrm{eff}}$. The ordering is monotonic, from
    $\langle r\rangle=0.39$ (smallest) to $0.55$ (largest). Open
    symbols are the independent three-species run; they fall on the
    same curve. The sorting is a property of the mechanism, not of the
    number of species used to sample it.
    (\textit{b}) Adjacent-pair separation $\Delta\langle r\rangle$
    against $\Delta(1/\Pe_{\mathrm{eff}})$, prediction~3 of
    \S\ref{sec:front}. The fit is constrained through the origin
    because zero contrast must give zero separation identically; the
    origin is not a free parameter. The through-origin slope is
    $\approx1.6\times10^{3}$ ($1.58$--$1.66\times10^{3}$ across the
    precursor-cutoff band, $R_{\mathrm{cut}}=1.05$ against the full
    mesh, about $5\%$), the common slope $|\Delta^{(1)}|$ of
    \S\ref{sec:expansion}, with $R^2_0=0.996$ ($R^2_c=0.989$ against
    the mean). The ordering is insensitive to the cutoff; the slope
    is not. $\Pe_{\mathrm{eff}}$ values are those of the five-species
    run and differ slightly from the three-species values quoted in
    figures~\ref{fig:enrichment} and \ref{fig:dynamics}, because the
    diffusion floor is set per run (\S\ref{sec:gate}).}
  \label{fig:scaling}
\end{figure}

Once the common-front premise holds the deposit ordering becomes a statement
about a single scalar per species. Figure~\ref{fig:scaling}(a) plots the mean
deposit radius $\langle r \rangle_i$ against $1/\Pe_{\mathrm{eff},i}$ for
the five-species run. The ordering is monotonic, from $\langle r \rangle = 0.39$
for the smallest species to $0.55$ for the largest. The open symbols from the
independent three-species run fall on the same curve. So the sorting is a
property of the mechanism, not of how finely the size distribution is sampled.
Three species and five species trace one relation, because both read out the
same underlying map from diffusivity contrast to radial position.

The band separation is the quantitative core of the result.
Figure~\ref{fig:scaling}(b) plots the adjacent-pair separation $\Delta\langle r
\rangle$ against the inverse-P\'eclet contrast
$\Delta(1/\Pe_{\mathrm{eff}})$, the relation pre-registered as
Prediction~3. The through-origin fit gives a slope of
$\approx 1.6\times10^3$ ($1.58$--$1.66\times10^3$ across the
precursor-cutoff band), the common slope $|\Delta^{(1)}|$ of
equation~\eqref{eq:meanr}, with $R^2_0 = 0.996$ ($R^2_c = 0.989$
against the mean). Separation scales linearly with diffusivity
contrast at fixed sweep history, and at the dry-out readout the
linear law holds across the full measured contrast range. The sag of
the largest-contrast pair below the line, visible at later snapshots,
is post-dry-out smearing (\S\ref{sec:deposit}) rather than resolved
$O(D_i^2)$ saturation; the expansion's neglected $O(D_i^2)$ term
bounds any true departure from linearity, and within the present
resolution none is detected. We read this
slope as the resolution of the contact-line chromatograph: the radial
separation delivered per unit of diffusivity contrast, and the quantity a
fractionation application would tune. The shaded band spans the sensitivity of
the slope to the precursor cutoff ($R_{\mathrm{cut}} = 1.05$ against the full
mesh, about $5\%$).

The two channels do not respond to that cutoff in the same way. The direction
of the ordering is insensitive to it. The slope is not. That split is not
incidental; it is the \S\ref{sec:theorem} structure made visible. Direction is
fixed by the size-symmetry of the operator, so it survives any modelling choice
that preserves that symmetry. Magnitude is model-dependent, and the flow
closure, the rheology, or the cutoff can each move it. The resolution law is
real, and it is measured here, but we report it as a leading-order scaling with
a stated sensitivity band, not as a closure-independent constant.

\section{Discussion and conclusions}
\label{sec:discussion}

The reversal to a small-rich centre and large-rich annulus is not
unique to the substrate-free geometry. \citet{iqbal2018} report the
same direction for a bidisperse droplet on a hydrophobic solid
($\theta\approx110^\circ$): once the contact line depins, an inward
surface-tension force carries both species toward the apex and strands
the large species in an intermediate ring. That mechanism is a force
balance (surface tension against van der Waals, electrostatic and drag
forces), resolved size by size, with diffusion absent. Their
geometry is a blunt high-angle cap with no wedge and no receding-front
boundary layer; the curvature is nowhere singular. Ours is diffusivity
contrast read out in a boundary layer at a low-angle receding front, on
no solid at all. The two systems share a sorting direction and nothing
else.

This reading also fixes where the effect must vanish. \citet{jain2025}
solve the bidisperse sessile droplet by VOF--DEM and find that with
Marangoni stresses absent, the receding-contact-line flow sweeps both
species to the apex and produces no size sorting; sorting appears only
when thermocapillary stress is switched on. Their particles are
micron-scale: $\Pe$ is enormous and the diffusivity contrast between
species is negligible. That is exactly the limit in which the present
mechanism predicts no sorting. Their null result is the
$\Pe\to\infty$ corner of this model, confirmed by an independent full
solve. The contrast that drives fractionation here lives in the
few-nanometre regime, and there the receding front sorts without
Marangoni stresses and without a force balance.

Size enters the transport problem in one place, the Stokes--Einstein
diffusivity, and nowhere else. With the species operator identical across all
species but for a scalar P\'eclet number, the ordering is fixed by the structure
of the problem: a small-rich centre and a large-rich annulus is the only
arrangement the mechanism can produce. Marangoni stresses, colloidal forces and
wetting details do not set the direction. They set the magnitude, which is
model-dependent and carries the application: adjacent bands separate in linear
proportion to their inverse-P\'eclet contrast, a resolution law whose slope is
the resolution of the contact-line chromatograph and the quantity a
fractionation application would tune.
What that buys is worth stating concretely. For a pair of neighbouring
sizes at $\hat a=2.0$ and $2.5$~nm, near $\Pe\sim10^4$, the
inverse-P\'eclet contrast is $\Delta(1/\Pe)\approx
2\times10^{-5}$, so the measured slope predicts a band separation
$\Delta\langle r\rangle\approx0.03$: at $\hat L_0\sim1$~mm, some
$30\,\mu$m of radial separation for a $25\%$ difference in radius,
which is resolvable by ordinary optical or small-angle scattering
mapping of the dried film.
The ordering also runs opposite to the
coffee ring; removing the solid and the pinned line inverts the direction the
pinned-substrate literature had settled on, as an output of the substrate-free
geometry rather than a modelling choice. The reversal is strongest in the
few-nanometre regime, and that is where it matters. There transport is
diffusive, sedimentation and filtration lose their purchase, and the
inverse-P\'eclet contrast between neighbouring sizes is steepest. So the
front sorts most strongly in exactly the size range that is hardest to
sort by any other means: no external field, no functionalised surface.
Whether the effect
survives in a real evaporating film is an experimental question, and a
well-posed one.

\subsection{Regime of validity}
\label{sec:regime}

\begin{figure}[t]
  \centering
  \safeincludegraphics[width=0.6\linewidth]{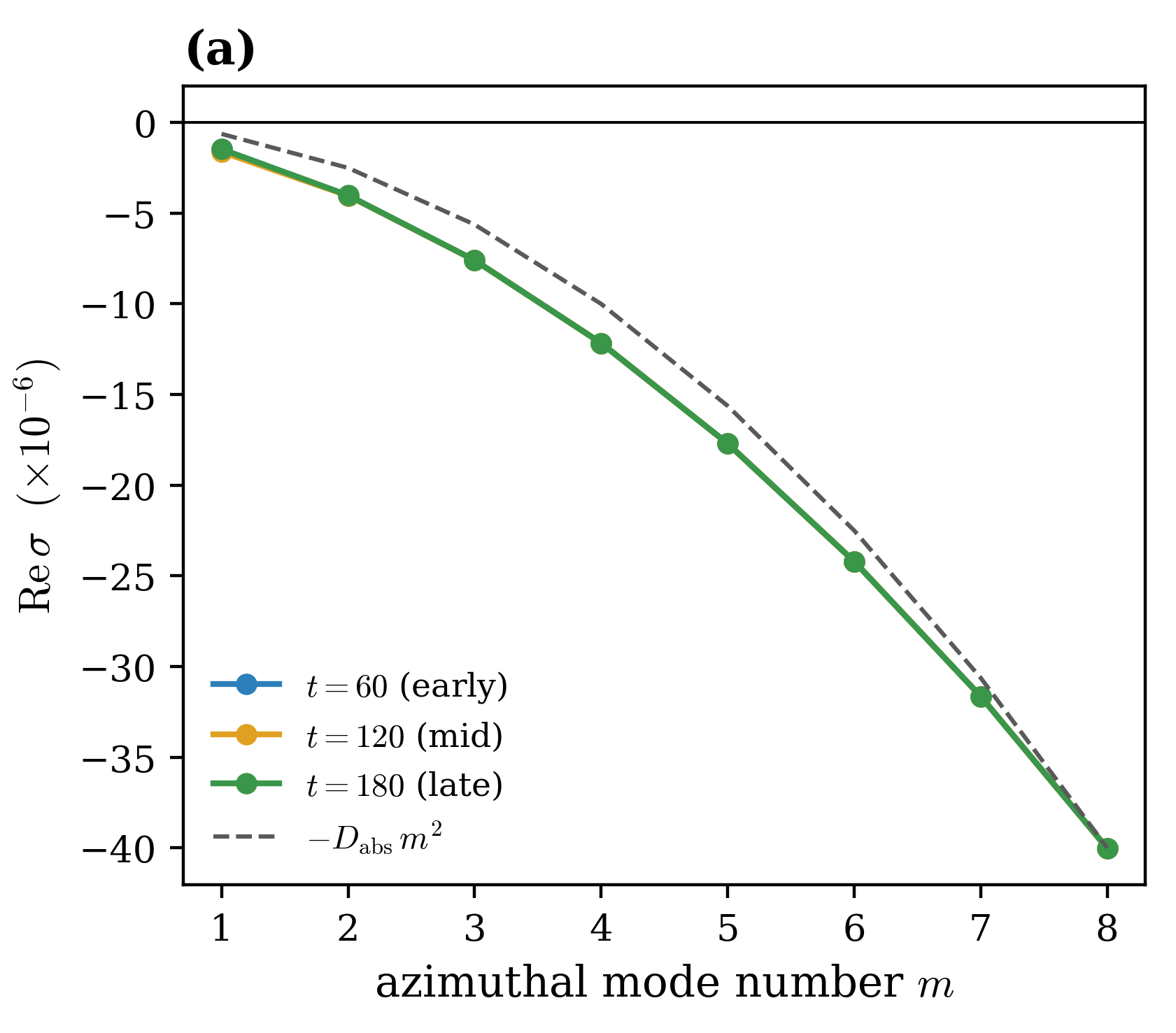}
  \caption{Azimuthal linear stability of the loaded receding front.
    Growth rate $\mathrm{Re}\,\sigma(m)$ of azimuthal normal modes
    $m=1$--$8$ about the axisymmetric base state at three snapshots
    spanning the deposition window ($t=60,\,120,\,180$; dry-out at
    $t_{\mathrm{dry}}\approx189$), perturbing the full state including
    the subphase deflection $s$. All modes decay at all snapshots:
    $\mathrm{Re}\,\sigma<0$ throughout, the three snapshot curves
    collapse, and damping deepens monotonically with $m$. The leading
    rate is $\omega=\max_{t,m}\mathrm{Re}\,\sigma=-1.5\times10^{-6}$,
    giving $|\omega|\,t_{\mathrm{dry}}\approx3\times10^{-4}\ll1$: over
    a drying time the azimuthal modes neither grow nor decay
    appreciably. The dashed line marks the numerical diffusion floor
    $-D_{\mathrm{abs}}m^2$, which the damping magnitudes track, so the
    measured rates are a bound rather than a resolved physical decay;
    the conclusion drawn is correspondingly that no azimuthal growth is
    detectable above that floor over the deposition window, which is
    what the axisymmetric reading of the deposit requires. It rests on
    $|\omega|\,t_{\mathrm{dry}}\ll1$, not on the magnitude of the
    damping.}
  \label{fig:azimuthal}
\end{figure}

The kinetic closure \eqref{eq:J} is the one modelling element that a
room-temperature realization would replace outright. Decane at
ambient conditions is vapour-diffusion-limited, and the generalized
closure of \citet{sultan2005} carries a non-local flux profile that
\eqref{eq:J} does not. That refinement acts on the front kinematics
$\dot r_f$, and through them on the spike widths \eqref{eq:delta} and
the resolution slope of figure~\ref{fig:scaling}. It is a magnitude
revision and we expect it to be a substantial one. It cannot move the
ordering, which by \S\ref{sec:expansion} requires only that the front
recede monotonically inward, a property the edge-peaked
diffusion-limited flux reinforces rather than threatens.

\begin{equation}
\eps\ll1;\qquad
\mathrm{Re}\,\eps\ll1;\qquad
\hat\mu_d/\hat\mu_e\approx0.05\ll1;\qquad
\hat S>0;\qquad
\eps^2\Pe_i\ll1\ll\Pe_i;\qquad
\tau\ll1;
\end{equation}
isothermal baseline; kinetic evaporation closure (see above);
dilute-to-moderate loading with collective
rheology;
axisymmetry \emph{tested}. The claims split under azimuthal
perturbation. The $\langle r\rangle$--$\Pe$ ordering of
figure~\ref{fig:scaling} is an azimuthal average of a radially driven
mechanism and survives front corrugation: a fingered front still sorts
by diffusivity contrast along every radial cut. The ringed morphology
of figure~\ref{fig:deposit3} is the axisymmetric realization and is
contingent on stability of the loaded front on the compliant subphase
(cf.\ the symmetry-breaking of \citealt{malachtari2025} on soft
solids).
A linear stability analysis of azimuthal normal modes $m=1$--$8$ about
the axisymmetric base state, perturbing the full state including the
subphase deflection $s$, finds no azimuthal growth at any snapshot
across the deposition window (figure~\ref{fig:azimuthal}): every mode
decays, at rates that track the numerical diffusion floor, so what the
analysis establishes is a bound---no corrugation develops above that
floor over a drying time---rather than a measured decay rate. On that
basis the ringed morphology is the realized state, not an idealization.
The contrast with the symmetry-breaking of \citet{malachtari2025} is
physical rather than incidental. Their compliant substrate is an
\emph{elastic} solid that stores shear stress capable of feeding a
corrugation; the present liquid subphase is restored by buoyancy and
interfacial tension alone; it relaxes rather than stores.
The composition-ordering argument of \S\ref{sec:theorem} is structural
and survives any refinement of the flow model that preserves the
species-symmetry of the operator; full ALE/Navier--Stokes simulation is
therefore a magnitude refinement, deferred.

\subsection*{Acknowledgements}
The author is grateful to George Karapetsas for correspondence on the
thin-film formulation and for discussions of the model derivation, to
Tobias Hanrath for his guidance and support throughout this work, and
to the members of the Hanrath group for helpful discussions during its
early stages.
The author acknowledges the use of Claude (Anthropic), principally
Claude Opus 4.6, accessed via the claude.ai web interface between
April and July 2026, for language editing and stylistic refinement of
the manuscript text, and for independent checking of the analytical
derivations of \S\ref{sec:expansion} and
Appendix~\ref{sec:depthavg}.
Earlier model versions were used for the same purposes during the
preparation of preliminary drafts. No text was incorporated without
author revision, and no content, data or references were generated by
the tool for direct inclusion. All modelling decisions, the
mathematical formulation and the analysis are the author's own, and
the author takes full responsibility for the content.

\subsection*{Funding}
The author acknowledges funding support from Cornell University.

\subsection*{Competing interests}
The author declares none.

\subsection*{Data availability statement}
The simulation code (a pyoomph implementation of the model of
\S\ref{sec:formulation}), the run configurations for every case
reported, the data underlying all figures, the gate-off control of
\S\ref{sec:gate}, and the moment-rate decomposition of
\S\ref{sec:expansion} are openly available at
\url{https://doi.org/10.5281/zenodo.21420067}.

\appendix

\section{The subphase}
\label{sec:subphase}

\subsection{Quasi-static balance: Winkler-by-Archimedes}
\label{sec:winkler}

The subphase is deep and viscous. The normal-load balance at the
decane--EG interface reads, at leading order,
\begin{equation}
\underbrace{\hat\rho_e\hat g\,\hat s}_{\text{buoyant restoring}}
\;-\;\underbrace{\hat\gamma_{de}\,\hat\nabla^2\hat s}_{\text{interfacial tension}}
\;=\;-\Big(\underbrace{\hat\rho_d\hat g\,\hat h}_{\text{film weight}}
 \;+\;\underbrace{\hat p_{\mathrm{cap}}}_{\text{capillary load}}\Big).
\end{equation}
Nondimensionalizing with the \S\ref{sec:scaling} scales:
\begin{equation}
k_s\,s\;-\;\Cb\,\nabla^2 s\;+\;\Bo\,h\;+\;\big(p-\Pi\big)\;=\;0,
\qquad
k_s=\frac{\Bo}{\Rrho}.
\tag{S}
\label{eq:S}
\end{equation}

Three remarks. \textbf{(a)} The stiffness $k_s=\Bo/\Rrho$ is fixed by
Archimedes, not chosen: neglecting tension and capillary load,
\eqref{eq:S} gives $s=-\Rrho h$, isostatic flotation, which is also the
initial condition. \textbf{(b) The load is $p-\Pi$,
not $p$.} $\Pi$ is the excess film pressure arising from the direct
interaction between the two media bounding the film. The traction it
exerts on the lower interface is accompanied by an equal and opposite
direct gas--EG interaction traction transmitted across the film, so the
\emph{net} normal load delivered to the subphase is $p-\Pi$. This is
consistent with the scale separation (in the precursor, $\Pi$ sits at
the Kelvin value $-1/\Delta=-10^3$ purely as the volatility-suppression
device of \S\ref{sec:evap}, at a film thickness
$h_p\sim10^{-3}\hat H_0\sim\text{\AA}$ far below the continuum scale of
the interface-deflection problem). The numerics confirm it:
transmitting the full $p$ to the subphase demands
$s\sim+1/(\Delta k_s)\sim10^4$, four orders of magnitude of spurious
deflection under the precursor, and the first time step fails
($\mathrm{d}t\to10^{-12}$, verified). With $p-\Pi$ both subphase modes
integrate cleanly and $s$ stays within its physical range
$[-\Rrho\,h_{\max},\,O(0.1)]$.
\textbf{(c)} With $\Bo=0$ the subphase
decouples from $\Pdrv$ entirely; a rigid-mode smoke test therefore
validates nothing about \eqref{eq:S}. Recorded here because this is
exactly how the load bug survived initial testing.

\subsection{Relaxational embedding (code default, \code{SUB\_CONS}~$=0$)}
\label{sec:relax}

The code solves \eqref{eq:S} by pseudo-time relaxation,
\begin{equation}
\tau\,\partial_t s=\Cb\,\nabla^2 s
 -\big(k_s s+\Bo\,h+(p-\Pi)\big),
\tag{S$_\tau$}
\label{eq:Stau}
\end{equation}
which has \eqref{eq:S} as its fixed point. The physical subphase
relaxation time (capillary levelling of a deep viscous layer,
$\hat t_{\mathrm{rel}}\sim\hat\mu_e\hat\lambda/\hat\gamma_{de}\sim
0.1$--$1.4$~ms for $\hat\lambda\sim0.1$--$1$~mm) is
$10^{-4}$--$10^{-2}$ of the drying time, so the physical regime
\textbf{is} the quasi-static limit. Equation \eqref{eq:Stau} with
$\tau\le10^{-1}$ is then simply a solver for that limit and makes no
claim to model EG hydrodynamics.
Validation: the dried deposits are insensitive to the pseudo-time
constant over
$\tau\in\{3\times10^{-3},10^{-2},3\times10^{-2},10^{-1}\}$, agreeing
to within $10^{-4}$ in the per-species mean deposit radius
$\langle r\rangle_i$ (maximum deviation $9.1\times10^{-5}$ across all
$\tau$ pairs; the pointwise deposit densities $q_i$ agree to
$4\times10^{-4}$), which certifies convergence to the quasi-static
limit and retires the anchoring question. The baseline
$\tau=10^{-2}$ is the value used for the dried deposits of
\S\ref{sec:results}. The known
spectral defect of \eqref{eq:Stau} (short wavelengths relax fastest,
inverted relative to a real layer) is immaterial \emph{in the
quasi-static limit}: only the fixed point survives.

\subsection{Conserved alternative (\code{SUB\_CONS}~$=1$)}
\label{sec:conserved}

To test whether finite-time subphase hydrodynamics (displaced-volume
transport; the EG dimple-and-rim) affects the deposit, the flux form
\begin{equation}
\tau\,\partial_t s
 =\frac{1}{r}\partial_r\!\left[r\,M_s\,
 \partial_r\!\big(k_s s+\Bo\,h+(p-\Pi)\big)\right]
 +\Cb\,\frac{1}{r}\partial_r\!\big(r\,\partial_r s\big)
\tag{S$_c$}
\label{eq:Sc}
\end{equation}
conserves $\int s\,r\dd r$ up to boundary exchange with the far-field
reservoir ($s=0$ at $r=L_x$), relaxes long wavelengths slowest, and
produces the rim ($s_{\max}>0$ outboard of the lens; observed,
$s_{\max}=+0.09$ at $t=2$): the liquid-subphase analogue of the
secondary wetting ridge of \citet{malachtari2025}. The
interfacial-tension term enters as conservative smoothing rather than as
fourth-order curvature inside the flux potential (which would require an
auxiliary field); at the front scale the $k_s$ restoring term dominates
and the distinction is higher-order.
Equation \eqref{eq:Sc} has not been carried to dry-out: the flux form
is numerically stiff and the integration fails well before the deposit
is written. The comparison it was designed for is therefore not
available, and the quasi-static anchoring of the results rests on the
$\tau$-sweep of Appendix~\ref{sec:relax}, which certifies the fixed
point of \eqref{eq:S} directly.
Whichever form is used, the subphase enters the species problem only
through the collective fields $h$ and $\ubar$, so by
\S\ref{sec:theorem} it can move the magnitude of the separation and not
its sign; what \eqref{eq:Sc} is designed to probe is a magnitude
question.
The finite-time question \eqref{eq:Sc}
poses, whether displaced-volume transport in the subphase alters the
deposit, remains open; disagreement, if found, would be discovered
physics and would escalate to the two-layer coupling of
\citet{karapetsas2011}.

\section{Conservation identities (numerical acceptance tests)}
\label{sec:conservation}

\begin{enumerate}
\item \textbf{Species (exact):}
$\dfrac{\dd}{\dd t}\displaystyle\int_0^{L_x}q_i\,r\dd r=0$ for each $i$:
machine precision by construction; any drift is a bug.
\item \textbf{Liquid:}
$\dfrac{\dd}{\dd t}\displaystyle\int_0^{L_x}h\,r\dd r
 =-E\displaystyle\int_0^{L_x}J\,r\dd r$: evaporation is the only sink;
audit that the precursor contributes $J\approx0$ (identity \eqref{eq:K}
holding numerically: $p_{\mathrm{prec}}=-1/\Delta=-1000$, verified).
\item \textbf{Subphase (conserved mode only):}
$\dfrac{\dd}{\dd t}\displaystyle\int s\,r\dd r=$ boundary flux at $L_x$
only.
\end{enumerate}

\section{Numerical method and validation}
\label{sec:benchmark}

\subsection{Discretization and timestepping}
\label{sec:nummethod}

The coupled system \eqref{eq:F}, \eqref{eq:P}, \eqref{eq:Ti} and the
subphase equation \eqref{eq:Stau} is solved with the
finite-element framework pyoomph \citep{diddens2024}, built on
oomph-lib \citep{heil2006}. The fourth-order film problem is posed as
the mixed pair $\{h,p\}$ with $C^2$ elements (\S\ref{sec:pressure});
the species loads $q_i$ are prognostic fields on the same mesh, and the
weak form of \eqref{eq:Ti} telescopes, conserving $\int q_i\,r\dd r$ to
machine precision at every resolution tested
(Appendix~\ref{sec:conservation}). The domain $r\in[0,L_x]$, $L_x=4$,
is discretized with a uniform mesh comprised of $N_{\mathrm{el}}$
elements; production results use $N_{\mathrm{el}}=1000$.
Temporal resolution was checked on the final deposits: three-species
runs at \code{maxstep} $2.0$ and $0.1$, otherwise identical, agree at
$t=200$ to within $1.4\times10^{-3}$ in the per-species mean deposit
radius $\langle r\rangle_i$ (comparable to the spatial resolution of
Appendix~\ref{sec:convergence}, so temporal error is subdominant to
discretization). (The
adaptive controller pins the effective step to the output interval at
\code{maxstep}~$\ge0.5$, so $0.1$ is the first genuine refinement.
The pair is run at $N_{\mathrm{el}}=500$: temporal convergence
requires only that the two runs share a mesh, and the spatial
question is settled independently in \S\ref{sec:convergence}.)

Every routine was validated against the conservation
identities of Appendix~\ref{sec:conservation}, the mesh and
sensitivity studies of this appendix, and the published benchmark of
\citet{karapetsas2016}. The complete code is released at the
repository cited in the data availability statement.

\subsection{Mesh and precursor-floor sensitivity}
\label{sec:convergence}

Spatial convergence was assessed on the two-species baseline
configuration at the front frame ($t=150$), where the contact-line
spike is sharpest and most demanding on resolution, with runs carried
to $t=200$ (past dry-out). Deposit densities $q_i=hc_i$ and the local
composition $E=q_s/(q_s+q_l)$ were compared across
$N_{\mathrm{el}}=500$, $1000$, $2000$ (reference) and $4000$, the spike
characterized by its height and peak radius within the lens
($0.1<r<1.1$).

\begin{figure}[tp]
  \centering
  \safeincludegraphics[width=\textwidth]{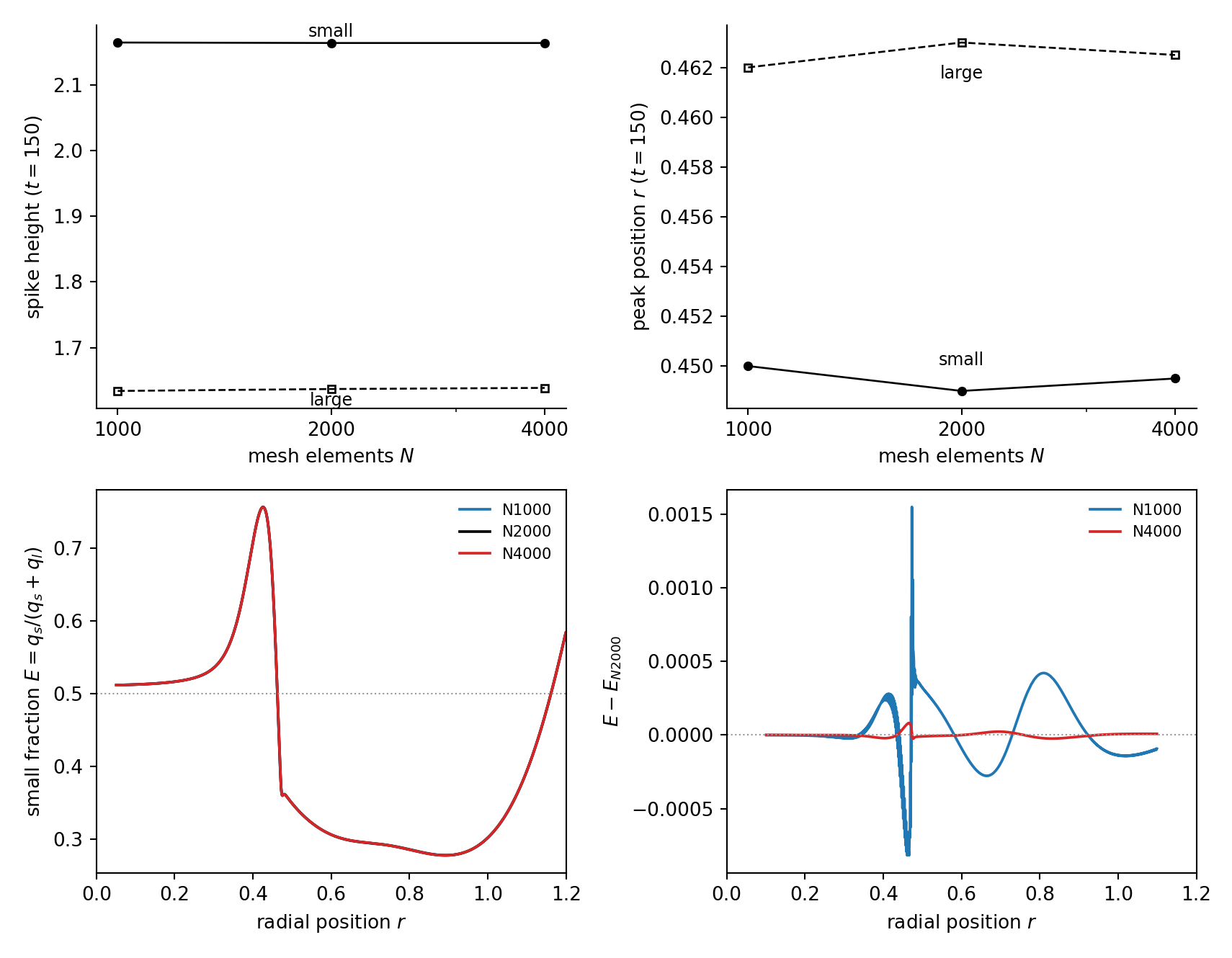}
  \caption{Mesh convergence of the front spike and deposit composition
    (two-species baseline, front frame $t=150$).
    \textit{Top:} spike height and peak radius for both species across
    $N_{\mathrm{el}}=1000$--$4000$: heights agree to better than
    $0.3\%$, radii to within $0.002$.
    \textit{Bottom left:} composition $E=q_s/(q_s+q_l)$ overlaid across
    resolutions; the curves are indistinguishable.
    \textit{Bottom right:} deviation from the $N_{\mathrm{el}}=2000$
    reference; the maximum deviation at $N_{\mathrm{el}}=1000$ is
    $1.5\times10^{-3}$, localized at the front discontinuity.}
  \label{fig:meshconv}
\end{figure}

For $N_{\mathrm{el}}\ge1000$ the study is converged
(figure~\ref{fig:meshconv}): spike heights agree to better than
$0.3\%$, peak radii to within $0.002$, and the composition profile at
$N_{\mathrm{el}}=1000$ deviates from the reference by at most
$1.5\times10^{-3}$ in $E$, localized at the front discontinuity. At
$N_{\mathrm{el}}=500$ the front \emph{position} and the species
\emph{ordering} are already correct (peak radii within $0.003$ of the
reference), but the amplitude is not: the large-species spike is
under-resolved by ${\sim}16\%$, and the under-resolved front imprints
spurious composition oscillations of amplitude ${\pm}0.03$ into the
deposit record as it sweeps. Since the deposit is the time-integrated
trace of the front (\S\ref{sec:frontal}), amplitude error at the front
is permanent in the deposit; production resolution is therefore set at
$N_{\mathrm{el}}=1000$. Consistent with the reporting standard used
throughout, the mesh study separates the two channels cleanly: the
sorting direction, ordering and crossover radius are grid-robust down
to the coarsest mesh tested; only amplitudes require the finer mesh.

\begin{figure}[tp]
  \centering
  \safeincludegraphics[width=\textwidth]{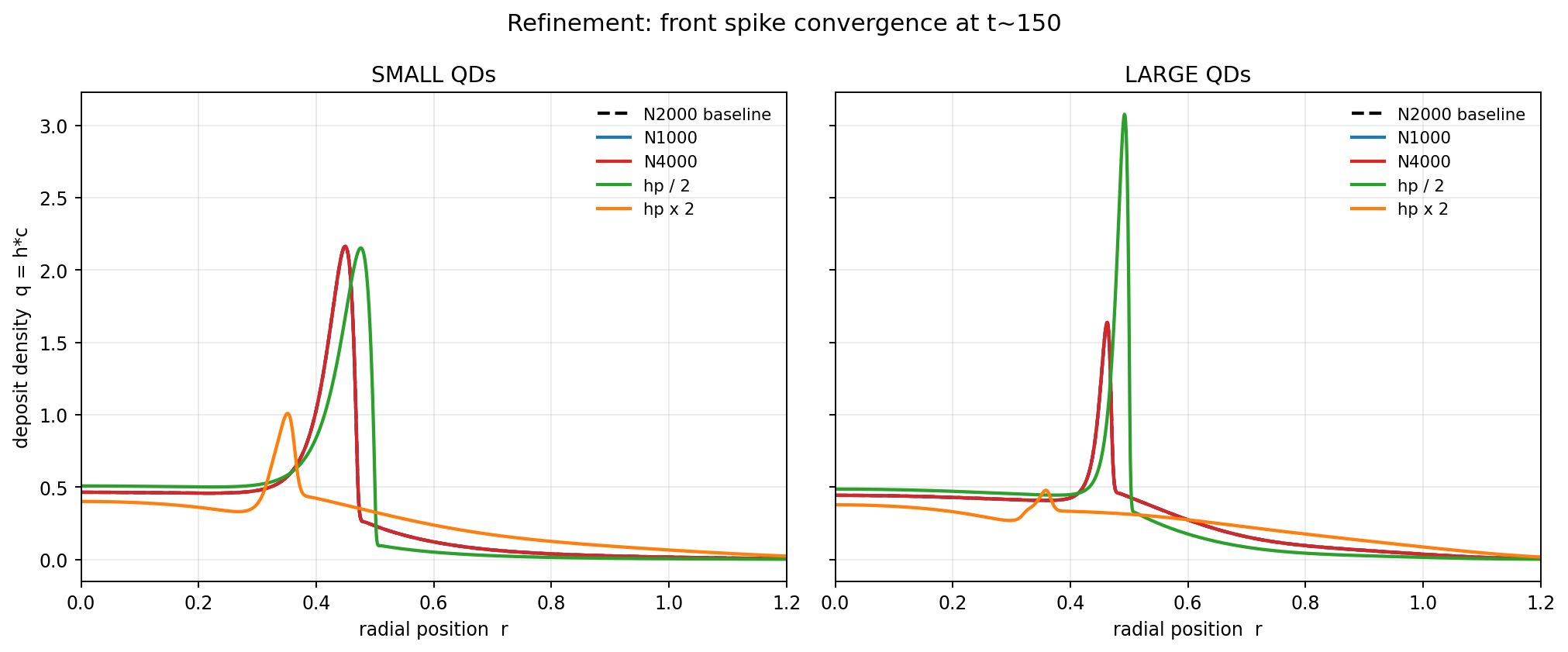}
  \caption{Front-spike sensitivity to mesh and precursor floor
    (two-species baseline, $t\approx150$): deposit density $q_i=hc_i$
    for the small (\textit{left}) and large (\textit{right}) species.
    The $N_{\mathrm{el}}=1000$, $2000$ and $4000$ curves coincide.
    Halving or doubling $h_p$ (with $A\propto h_p^3$ per
    \eqref{eq:K}) shifts the spike position and changes its
    amplitude, and the species' relative amplitudes, by $O(1)$
    factors, while the ordering (small spike inside large) is
    preserved at every floor value: direction robust, magnitude
    floor-sensitive.}
  \label{fig:floorconv}
\end{figure}

Sensitivity to the precursor thickness $h_p$ was probed by halving and
doubling it, moving $A$ as $A\propto h_p^3$ to preserve the identity
\eqref{eq:K} (figure~\ref{fig:floorconv}). The front spike responds
strongly: its position shifts with the effective dry-out radius and its
amplitude, including the \emph{relative} amplitudes of the two
species, varies by $O(1)$ factors. The structure protected by
\S\ref{sec:theorem} does not respond: at every floor value the
small-species spike sits inside the large-species spike. As $h_p$ is a
physical parameter of the
precursor model rather than a numerical control, this is a modelling
sensitivity, not a convergence failure; we treat the fractionation
direction as robust and its magnitude, to which $h_p$ contributes at
leading order, as model-dependent.

The refinement study was performed on the two-species baseline model.
The $N$-species production model shares its discretization exactly
(identical mesh construction, $C^2$ mixed $\{h,p\}$ elements,
identical domain) and its front physics: every species rides the same
free-surface boundary layer, which is the feature the refinement
certifies. The spatial resolution established here therefore transfers
to the production model, whose figures are computed at the certified
$N_{\mathrm{el}}=1000$.

\emph{Benchmark.} The benchmark protocol uses the same numerical
machinery with the evaporation closure of \citet{karapetsas2016}
($K+h$ resistance) and their base
parameters ($\eps=0.1$, $K=0.1$, $E=0.005$, $A=10^{-6}$,
$\Delta=10^{-3}$, $\gamma=0.1$,
$M_{\mathrm{surf}}=M_{\mathrm{par}}=\psi=\chi=0$). Front positions
match their reported values to within $2\%$ over the full drying
history ($x_f=0.64,\,0.48,\,0.36$ at $t=10,\,50,\,100$ against their
$0.63,\,0.48,\,0.36$), apex heights track their profiles including the
descent to $h(0)=0.998$ at $t=100$, and the reconstructed velocity
field reproduces the structure and magnitude of their surfactant-free
flow. The production model deliberately uses a different closure
(equation~\eqref{eq:J}); see \S\ref{sec:evap}.

\begin{figure}[tp]
  \centering
  \safeincludegraphics[width=\textwidth]{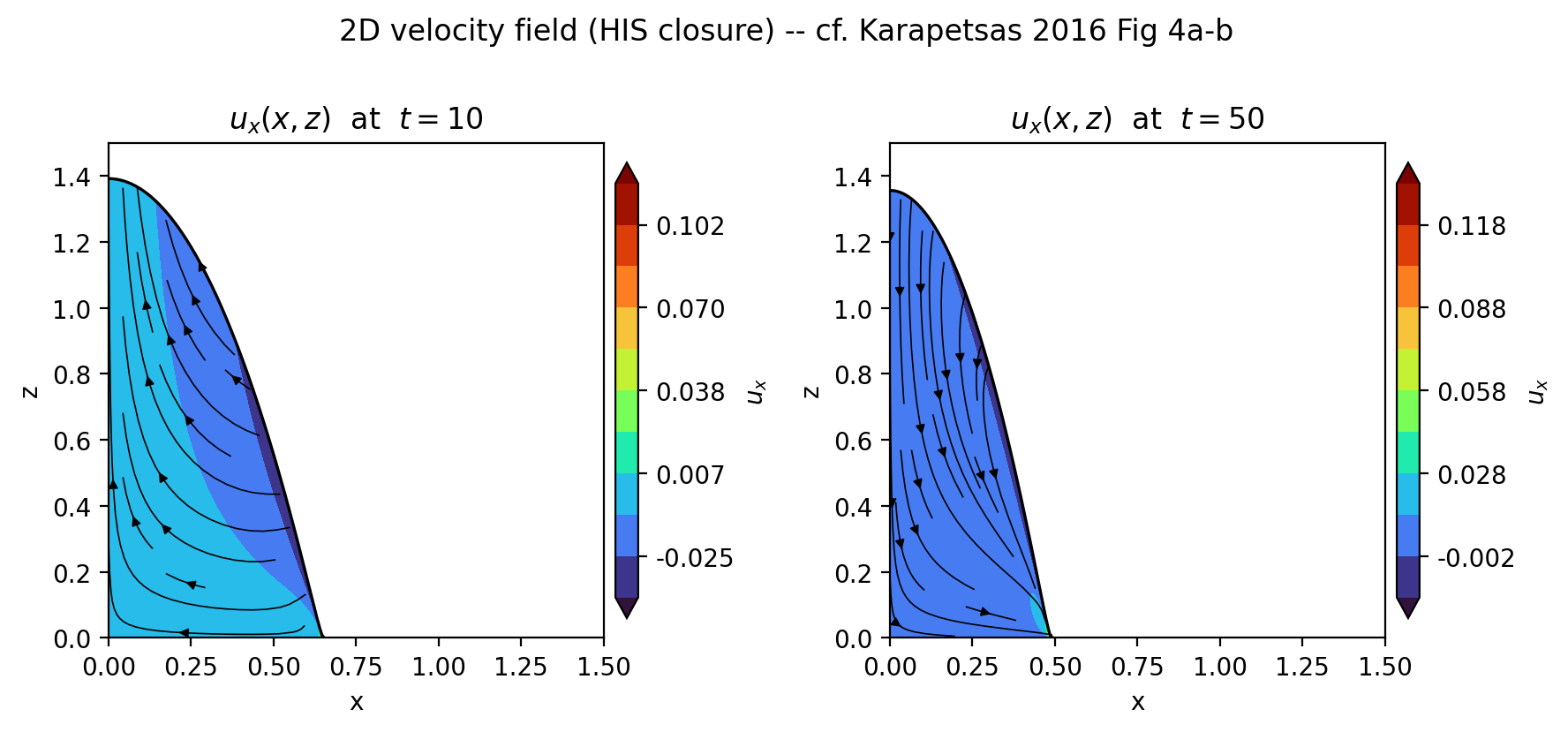}
  \caption{Validation against \citet{karapetsas2016}.
    Reconstructed velocity field $u_x(x,z)$ with streamlines at $t=10$
    and $t=50$ for the surfactant-free, particle-free case
    ($M_{\mathrm{surf}}=0$), computed with their evaporation closure and
    base parameters. Front positions match their reported values to
    within $2\%$ over the full drying history, and the flow field
    reproduces the structure and magnitude of their figure~4(a,b). The
    production model deliberately uses a different, conduction-in-series
    closure (equation~\eqref{eq:J}); see \S\ref{sec:evap}.}
  \label{fig:benchmark}
\end{figure}

\section{Depth-averaged species reduction}
\label{sec:depthavg}

Equation~\eqref{eq:AD} reduces to \eqref{eq:Ti} in two steps. Both use
only the no-flux conditions~\eqref{eq:noflux}.

\subsection{Vertical homogenization}
\label{sec:depthavg-vert}

Scaled, axisymmetric, the species equation \eqref{eq:AD} reads
\begin{equation}
\partial_t c_i+\frac1r\partial_r\big(r\,u\,c_i\big)+\partial_z\big(w\,c_i\big)
=\frac{1}{\Pe_i}\,\frac1r\partial_r\big(r\,\partial_r c_i\big)
+\frac{1}{\eps^2\Pe_i}\,\partial_z^2 c_i .
\label{eq:AD-scaled}
\end{equation}
The vertical diffusion carries the $\eps^{-2}$ enhancement. Expand
$c_i=\phi_i^{(0)}+\eps^2\Pe_i\,c_i^{(1)}+\dots$. At $O(\eps^{-2})$,
$\partial_z^2\phi_i^{(0)}=0$ with $\partial_z\phi_i^{(0)}=0$ at
$z=s,\eta$ (the $z$-projection of \eqref{eq:noflux}), so
\begin{equation}
\phi_i\equiv\phi_i^{(0)}=\phi_i(r,t).
\label{eq:zindep}
\end{equation}
Vertical structure is $O(\eps^2\Pe_i)$ and drops from the areal balance
below.

\subsection{Depth integration}
\label{sec:depthavg-int}

Write the total species flux $\bm J_i=c_i\bm u_d-D_i\hat\nabla c_i$, so
\eqref{eq:AD} is $\partial_t c_i+\hat\nabla\!\cdot\bm J_i=0$. Integrate
$\int_s^\eta\!\dd z$; Leibniz on each term, with
$q_i\equiv\int_s^\eta c_i\dd z$:
\begin{equation}
\partial_t q_i
+\frac1r\partial_r\!\Big(r\!\int_s^\eta\! J_i^{\,r}\dd z\Big)
-B_\eta+B_s=0,
\qquad
B_{z_b}\equiv\Big[\,c_i\,\partial_t z_b+J_i^{\,r}\,\partial_r z_b-J_i^{\,z}\,\Big]_{z=z_b}.
\label{eq:leibniz}
\end{equation}
Take $z_b=\eta$ with outward normal $\bm N_\eta=(-\partial_r\eta,\,1)$.
The kinematic identity for an interface point,
$\partial_t\eta+u_\Sigma^r\partial_r\eta=u_\Sigma^z$, gives
$\partial_t\eta+u\,\partial_r\eta-w=-(\bm u_d-\bm u_\Sigma)\cdot\bm N_\eta$,
and $\partial_r c_i\,\partial_r\eta-\partial_z c_i=-\hat\nabla c_i\cdot\bm N_\eta$.
Substituting $\bm J_i=c_i\bm u_d-D_i\hat\nabla c_i$,
\begin{equation}
B_\eta
=-c_i\,(\bm u_d-\bm u_\Sigma)\cdot\bm N_\eta
+D_i\,\hat\nabla c_i\cdot\bm N_\eta
=\big[\,D_i\hat\nabla c_i-c_i(\bm u_d-\bm u_\Sigma)\,\big]\cdot\bm N_\eta
=0
\label{eq:Beta}
\end{equation}
by \eqref{eq:noflux}. The bottom interface $z_b=s$ carries no phase
change, so $(\bm u_d-\bm u_\Sigma)\cdot\bm N_s=0$; the identical
manipulation gives $B_s=[\,D_i\hat\nabla c_i-c_i(\bm u_d-\bm u_\Sigma)\,]\cdot\bm N_s=0$,
again \eqref{eq:noflux}. Both boundary terms vanish identically.

What survives \eqref{eq:leibniz} is $\partial_t q_i+\frac1r\partial_r(r\!\int_s^\eta J_i^{\,r}\dd z)=0$.
With \eqref{eq:zindep} ($\phi_i$ off $z$), $\int_s^\eta u\dd z=Q=\ubar h$
and $\int_s^\eta\partial_r c_i\dd z=h\,\partial_r\phi_i$, so
\begin{equation}
\int_s^\eta\! J_i^{\,r}\dd z=\ubar\,h\,\phi_i-\frac{h}{\Pe_i}\,\partial_r\phi_i,
\qquad q_i=h\phi_i,
\end{equation}
and \eqref{eq:leibniz} collapses to \eqref{eq:Ti}.

\begin{remark}[no evaporative source]\label{rem:nosource}
The recession of $z=\eta$ enters only through $\bm u_\Sigma$ at the top,
inside the combination $(\bm u_d-\bm u_\Sigma)$ that \eqref{eq:noflux}
annihilates in \eqref{eq:Beta}. Evaporation therefore contributes no
source term to \eqref{eq:Ti}; it acts on composition only through $h$ in
$\phi_i=q_i/h$. Evolving $\phi_i$ directly reintroduces it by hand as
$(EJ/h)\phi_i$: Remark~\ref{rem:trap}.
\end{remark}

\section{Code $\leftrightarrow$ derivation dictionary}
\label{sec:dictionary}

\begin{center}
\small
\resizebox{\textwidth}{!}{%
\begin{tabular}{@{}llll@{}}
\toprule
Code & Derivation & Meaning & Status \\
\midrule
\code{h}, \code{zeta}, \code{p} & $h$, $s$, $p$ & film, subphase deflection, pressure & fields \\
\code{hc\%d} & $q_i=h\phi_i$ & areal species load (prognostic) & fields \\
\code{h\_pos}, \code{h\_reg} & $\tfrac12(h+\sqrt{h^2+4h_{\mathrm{reg}}^2})$, $h_{\mathrm{reg}}=h_p/10$ & smooth positive part & regularization \\
\code{Phi} & $\Pdrv=p+\Bo(h+s)$ & driving pressure & derived \\
\code{gam\_t} & $\Ct$ & scaled decane--air tension & fixes $\hat U$ \\
\code{gam\_b} & $\Cb=(\hat\gamma_{de}/\hat\gamma_{da})\Ct$ & scaled decane--EG tension & \textbf{anchored} ($11.7/23.8$) \\
\code{Rrho} & $\Rrho=\hat\rho_d/\hat\rho_e$ & density ratio & \textbf{physical} ($0.656$) \\
\code{k\_sub} & $k_s=\Bo/\Rrho$ & Archimedean restoring & \textbf{derived} \\
\code{A}, \code{Delta}, \code{hp} & $A$, $\Delta$, $h_p=(A\Delta)^{1/3}$ & Hamaker, Kelvin, precursor & \textbf{one identity (K)} \\
\code{K}, \code{E} & $K$, $E$ & resistance ratio, evaporation number & closure (J); RT pending \\
\code{PE[i]} & $\Pe_i\propto\hat a_i$ & per-species P\'eclet & Stokes--Einstein \\
\code{D\_REL}, \code{D\_ABS} & $\Pe_{i,\mathrm{eff}}=1/D_i$, $D_{\mathrm{abs}}=0.1/\max_i\Pe_i$ & numerical floor on $1/\Pe_i$ & \emph{per run}; keep $\ll$ physical $D$; \S\ref{sec:gate} \\
\code{LOAD\_K} & $k_L$ & loading-profile steepness & IC, \S\ref{sec:icbc}; production $k_L{=}10$, control $30$ \\
\code{tau\_sub}, \code{M\_SUB} & $\tau$, $M_s$ & pseudo-time / subphase mobility & quasi-static; $\tau$-sweep certified, App.~\ref{sec:relax} \\
\code{c\_jam} & $c_{\mathrm{jam}}=50$ & jamming threshold on $\ctot$ & size-blind; \emph{active}, gate-off control \S\ref{sec:gate} \\
\code{chi}, \code{PHI\_MAX} & $\chi$, $\Phi_{\max}$ & Krieger--Dougherty regularization & size-blind; direction-neutral \\
\code{S\_PIN}, \code{MAR}, \code{EVAP\_A} & n/a & ladder toggles & off at baseline by \S\ref{sec:theorem}; \code{EVAP\_A} swept, \S\ref{sec:evap} \\
\bottomrule
\end{tabular}}
\end{center}

\bibliographystyle{abbrvnat}
\bibliography{references}

\end{document}